\documentclass[mnsc,nonblindrev]{informs3_hide} 

\OneAndAHalfSpacedXI

\usepackage{natbib}
 \bibpunct[, ]{(}{)}{,}{a}{}{,}%
 \def\bibfont{\small}%

\newcommand\blfootnote[1]{%
  \begingroup
  \renewcommand\thefootnote{}\footnote{#1}%
  \addtocounter{footnote}{-1}%
  \endgroup

\newcommand{\comment}[1]{\textcolor{red}{[SH: #1]}}
}
\newcommand{\mathbbm}[1]{\text{\usefont{U}{bbm}{m}{n}#1}} 

\TheoremsNumberedThrough     
\ECRepeatTheorems

\EquationsNumberedThrough    

\MANUSCRIPTNO{}
\usepackage{bm}
\usepackage{wrapfig,lipsum}
\usepackage{graphicx}
\usepackage{booktabs}
\usepackage{algorithm}
\usepackage{algpseudocode}

\definecolor{tencent_blue}{RGB}{0, 82, 217}
\definecolor{tencent_orange}{RGB}{238, 126, 71}

\usepackage{subcaption}

\newcommand{\bbE}{\mathbb{E}}
\newcommand{\bbX}{\mathbb{X}}

\newcommand{\bone}{\mathbbm{1}}
\newcommand{\independent}{\perp \!\!\! \perp}

\newcommand{\bb}[1]{\left[#1\right]}

\newcommand{\bc}[1]{\left\{#1\right\}}

\usepackage{xcolor}

\begin{document}



\RUNTITLE{Extrapolation}

\TITLE{Enhancing External Validity of Experiments with Ongoing Sampling}

%


\ARTICLEAUTHORS{%
\AUTHOR{Chen Wang}
\AFF{The University of Hong Kong, \EMAIL{annacwang@connect.hku.hk}} 
\AUTHOR{Shan Huang*}
\AFF{The University of Hong Kong, \EMAIL{shanhh@hku.hk}} 
\AUTHOR{Shichao Han}
\AFF{Tencent Inc., \EMAIL{shichaohan@tencent.com}}
} 

\ABSTRACT{
Subjects in online experiments typically arrive and participate sequentially, which can compromise the external validity of experimental results due to temporal shifts in sample characteristics. This issue is particularly pronounced in A/B tests, which are often conducted over short durations to support rapid product decisions. 
To address this challenge, we introduce a novel framework that adapts to the dynamic nature of participant arrivals to improve external validity in online experimentation. Our method segments the ongoing sampling process into three distinct stages---unstable, overlapping, and representative---each corresponding to different levels of generalizability. Leveraging survival analysis, we develop a heuristic function that detects these stages in real time and enables the construction of stage-specific estimators for the population average treatment effect (PATE).
We evaluate our framework using both a real-world A/B test and a platform-scale application involving 600 experiments conducted on WeChat, a major social media platform. The results show that our approach increases true positive decision rates by 28--37\% while reducing false positive rates by 17--29\%. We further provide evidence that the framework significantly influences experimentation practices—such as stopping decisions and experiment duration—following deployment, and we offer practical guidelines for implementation.

}


\maketitle
\blfootnote{* To whom correspondence should be addressed.}
\section{Introduction}
\label{sec:intro}
Randomized controlled experiments, which estimate treatment effects relative to a control condition, are widely regarded as the gold standard for causal inference. A/B tests, as online randomized controlled experiments, are extensively conducted in the technology industry to inform daily product management, including high-stakes, multi-million-dollar decisions such as the rollout of new features or promotional campaigns~\citep{kohavi2020trustworthy, feit2019test}. As \citet{shulman2023marketing} highlights, the marketing community has significant potential to contribute meaningfully to product management, where A/B testing has become a standard tool.
Researchers also increasingly leverage such experiments for scientific discovery.
The random assignment of participants to experimental groups generally ensures strong internal validity, enabling unbiased estimation of the sample average treatment effect (SATE). However, this methodological rigor does not inherently address the challenge of external validity—that is, the extent to which experimental results generalize to the broader population targeted by the intervention~\citep{rothwell2005external}.

Unlike traditional experiments, online experiments typically employ an ongoing sampling process, in which participants are continuously enrolled as they interact with the platform, rather than being recruited in a single pre-experiment phase \citep{kohavi2012trustworthy}. In this ongoing process, users encounter experimental treatments through their voluntary engagement with the platform; thus, the timing of their inclusion depends on when they arrive, and experimenters have no control over whether or when users participate. Consequently, the sample composition is inherently affected by the specific time period and duration of experiment, potentially compromising sample representativeness and results generalizability and raising concerns about external validity.

For instance, participants who join experiments earlier tend to be more active users and may respond differently to the treatment than those who enroll later---a phenomenon known as ``heavy-user bias'' \citep{wang2019heavy}. Short experiments can be particularly prone to this bias, as they oversample heavy users and thus yield biased estimates of treatment effects. Furthermore, restricting sampling to a short time window increases the risk of capturing specific user subgroups due to the ``periodic effect.'' Users tend to interact with a product---and potentially enter an experiment---according to schedules shaped by their characteristics \citep{finney1950example}. For instance, experiments conducted over a weekend will disproportionately include users who do not work on weekends, relative to weekday experiments.

While continuous enrollment enables rapid experiment scalability and product iteration, it complicates efforts to distinguish true treatment effects from artifacts of sampling variability over time. 
Importantly, the target population is experiment-specific. Each test is intended to estimate the treatment effect for a predetermined set of users, and different experiments may legitimately target different groups. Yet the realized experimental sample shaped by who arrives and participates during the test window may not match the target population in its characteristics, inducing estimation bias for the target estimand. For example, if an A/B test indicates that a new feature enhances user engagement but the results cannot be generalized to the target user base, the feature may fail to deliver the expected performance upon launch. Current experimentation practices often focus on achieving sufficient statistical power by recruiting a large enough sample—a goal that can typically be met within a short time frame on large platforms. However, this approach does not necessarily address the issue of sample unrepresentativeness resulting from the ongoing sampling process~\citep{rothwell2005external}.

While intuition suggests that extending an experiment’s duration might help align the sample more closely with the population, practical constraints render this approach challenging. Prolonged experiments impose substantial costs on firms, including risks to user experience (e.g., prolonged exposure to suboptimal features) and delays in product iteration cycles \citep{huang2023estimating}. Moreover, continuously monitoring the discrepancy between sample and population distributions for each experiment in real time is costly and impractical, as it requires substantial additional data storage and infrastructure investment.
In addition, comparing the two distributions in real time can further introduce sequential testing problems, which require significant computational capacity to address.

Existing methodological frameworks for generalizing experimental results—such as weighting, stratification, and transportability techniques \citep{stuart2011use, tipton2013improving, dahabreh2019generalizing, degtiar2023review}—assume a static experimental sample drawn from a well-defined population. For example, methods like inverse probability weighting \citep{dahabreh2020extending} depend on pre-specified population covariates, which become obsolete when the population evolves mid-experiment. Similarly, transportability frameworks \citep{egami2023elements} assume a fixed target population, making them unsuitable for addressing the temporal heterogeneity inherent to online platforms.

In this paper, we propose a framework that dynamically assesses sample representativeness and develops stage-specific estimators for the Population Average Treatment Effect (PATE), ensuring the generalizability of experimental results across varying experiment durations. Our framework explicitly addresses sample bias arising from non-representative sampling and provides systematic adjustments to mitigate its impact on treatment effect estimation.
By bridging the gap between experimental findings and real-world applicability, our approach enables product decisions to be grounded in evidence that more accurately reflects the broader target population. Rather than assuming that experimenters can continuously recover the evolving sampling process, our framework offers a cost-effective approach to delivering actionable, real-time guidance under partial observability and stochastic user arrivals.

Specifically, we begin by delineating three distinct stages in the ongoing sampling process of experiments: the unstable stage, the overlapping stage, and the representative stage. These stages are identified based on two specific criteria. The first criterion ensures that the probability of users with diverse covariates participating in the experiment is sufficiently high, enabling valid causal inferences to bridge the gap between the sample and the broader population. The second criterion verifies that the selected sample and the target population exhibit no significant differences across covariates. To identify these criteria, we develop a heuristic function based on the estimated probability of participation across covariates in real time. This function leverages survival analysis models. Compared to conventional probability estimation approaches (e.g., logistic regression), survival models excel in handling censored data---users who have not yet participated have unknown future participation times---and accommodating the ongoing process of sampling. 
Specific thresholds of the heuristic function are tied to the two criteria, enabling dynamic identification of the sampling stages. 

After identifying the stages, we introduce stage-specific strategies for estimating the average treatment effect on the target population. For the unstable stage, we highlight the inevitable bias in estimation and caution experimenters not to stop experiments during this stage. Reliable causal inferences cannot be made during this phase due to insufficient participation of users with diverse covariates. For the overlapping stage, we utilize the probability of participation estimated by the survival model to construct a bias-adjusted estimator, which rectifies the gap between the sample and the population. 
For the representative stage, we demonstrate that a simple difference-in-means estimator with lower variance can be directly employed to generate the average treatment effect estimation. At this point, the sample is sufficiently representative of the population, and no significant differences in covariates exist.

Finally, we evaluate our method and demonstrate its effectiveness using two levels of empirical evidence. 
First, we apply our framework to a real-world A/B test conducted on WeChat, a leading digital platform owned by Tencent. Like many major technology companies such as Google, Meta, and LinkedIn, Tencent conducts thousands of A/B tests for daily product management, and these experiments commonly face risks of sample unrepresentativeness under ongoing sampling. In this case study, we show that our framework recovers the true treatment effect as early as the overlapping stage and subsequently in the representative stage—two days earlier than the unadjusted estimator—thereby enabling faster yet reliable product decisions. 

Second, to demonstrate the scalability and practical applicability of our framework, we apply it to 600 online experiments conducted on WeChat. The results show that our method increases the true positive rate (TPR) by approximately 28--37\% while simultaneously reducing the false positive rate (FPR) by about 17--29\%. These findings indicate that our framework is able to identify more effective treatments without misclassifying a greater number of ineffective ones. Furthermore, 
as our framework has been deployed as an integrated function within WeChat's A/B testing system, we are able to observe its impact on experimenters' behavior, particularly in terms of stopping decisions and experiment duration. Comparing the 12-month periods before and after deployment, we find that experiments run significantly longer on average, with a higher proportion concluding in the overlapping and representative stages and fewer terminating in the unstable stage.

Managerially, our framework provides a transparent and operational tool for monitoring sample representativeness and improving the generalizability of experimental results in real-time online experiments. In online experimentation environments, the evolution of sample representativeness under ongoing sampling remains largely opaque to experimenters. Our approach opens this ``black box'' by offering a cost-effective and systematic way for platforms to track how the representativeness of the experimental sample evolves over time.

Beyond improving estimation accuracy, the framework supports more informed experimentation decisions by providing flexible guidance rather than imposing a single optimal stopping rule. Different stakeholders may prioritize different trade-offs between speed and reliability. For example, product managers who prioritize rapid iteration may choose to conclude experiments during the overlapping stage, accepting a modest reduction in decision quality in exchange for faster decisions. In contrast, data scientists seeking stronger evidence before endorsing a new product strategy may prefer to wait until the representative stage, where a simple estimator with lower variance, applied to a sample that more closely reflects the target population, ensures higher decision quality.

The framework also serves as a governance tool for experimentation processes. By identifying unstable stages in which reliable inference cannot yet be obtained, it helps prevent experiments from being terminated prematurely. In practice, experiments are sometimes stopped early—either to accelerate product deployment or due to organizational pressures in decision-making environments. By providing objective, real-time indicators of sample representativeness, our approach helps organizations mitigate such risks and improve the reliability of experimentation-based product decisions.

\noindent
\textbf{Roadmap.}
Section~\ref{sec:related_work} reviews related work. Section~\ref{sec:empirical} introduces a real-world experiment as a motivating example. Section~\ref{sec:probSetup} formalizes the three sampling stages, identification assumptions, and estimation methods. Section~\ref{sec:heuristics} develops the survival-analysis-based heuristic for stage determination. Section~\ref{sec:results} presents empirical results from a real-world experiment, and 600 platform experiments. Section~\ref{sec:practicalGuide} provides practical guidelines. Section~\ref{sec:conclusion} concludes.

\section{Related Work}
\label{sec:related_work}
Within the realm of external validity \citep{bracht1968external,campbell1986relabeling,findley2021external}, which determines the practical applicability of experimental findings, we address a specific aspect --- the extent to which experimental results can be generalized to a target population beyond the specific sample studied \citep{bell2016estimates,egami2021covariate,susukida2016assessing,tipton2016site,stuart2017generalizing,braslow2005generalizability,lesko2017generalizing}. 
While much attention has been given to extrapolating findings from static and fixed samples to populations, there has been limited exploration into assessing the representativeness and enhancing the generalizability of results from dynamic samples over time. \cite{egami2023elements} recognize time as an important contextual factor, arguing that findings are generalizable as long as the contextual effects can be fully captured by certain moderators. The heuristic function we establish serves a role similar to these moderators.
Using these connection variables, the treatment effect on populations of interest can be inferred through weighting, resampling, stratification, regression, and matching-based estimators \citep{andrews2017weighting,stuart2011use,tipton2013improving,dahabreh2019extending,kern2016assessing,dahabreh2020extending}. A key aspect of this process involves assessing several identification assumptions to ensure the unbiasedness of these estimators. For dynamic populations gathered through experiments, we address this challenge through heuristic methods.

A key element in generalizability analysis is quantifying the difference in participation between the sampled units and the target population. To address this, our paper utilizes survival analysis \citep{jenkins2005survival, klein2003survival, clark2003survival, liu2012survival, machin2006survival}, a branch of statistical methods designed for analyzing time-to-event data and widely applied across various domains \citep{demediuk2018player,hu2021personalized,hubbard2010gee,de1999mixture,kelly2000survival,aral2012identifying}. Survival analysis focuses on understanding the time until an event of interest occurs and identifying factors that influence the likelihood of the event happening at any given time.
We extend the application of survival models in two novel ways that go beyond their traditional use in clinical and reliability research. First, we repurpose the lifetime distribution function to estimate the time-varying probability of experiment participation conditional on covariates, which directly serves as the weighting mechanism for bias-corrected treatment effect estimation. Second, we leverage it to introduce heuristics that assist in identifying the stage criteria for experiments. This dual use of survival models is, to our knowledge, novel in the causal inference literature.

Recent research on determining the experimentation duration has focused primarily on ensuring internal validity - achieving an unbiased estimation of the causal effects on participants \citep{slack2001establishing,stuart2011use}. A common approach involves waiting until a sample of sufficient size is collected, thereby achieving the desired statistical power for the experiment \citep{simester2022sample,kohavi2020trustworthy,viechtbauer2015simple,lenth2001some}. Beyond fixed-sample experimentation, advancements in sequential testing allow experimenters to conclude discoveries and halt experiments at any point while maintaining a guaranteed false discovery rate (FDR) \citep{johari2022always,maharaj2023anytime,schonbrodt2017sequential,deng2016continuous}. Other studies explore early stopping of experiments by considering worst-case subpopulation effects to mitigate potentially large-scale negative consequences \citep{jeong2020assessing,adam2023should}. In contrast to these approaches, our work emphasizes ensuring the external generalizability of experimental results and the robustness of conclusions to the broader population, including both participants and non-participants —such as those informing product decision-making—derived from estimators based on dynamic samples. Instead of imposing a black-box optimal stopping point, we offer a white-box approach that enables diverse stakeholders to interpret sample representativeness during ongoing sampling and to construct estimators for the unbiased PATE at different stages of the experiment.

\section{Motivating Example}
\label{sec:empirical}
In this section, we present a real-world experiment conducted on WeChat that serves as a motivating example to introduce our framework and showcase its typical context.
In this experiment, WeChat examined the impact of a newly developed recommendation algorithm on its content search engine that displays a list of search results after users submit a query.\footnote{Due to a Non-Disclosure Agreement (NDA), the specific product line within the WeChat ecosystem where the algorithm was tested cannot be disclosed.}
Specifically, the treated algorithm ranks search results by prioritizing content that users have consumed in the previous week, while the control condition (status quo) does not apply this prioritization.
The primary outcome metric is the click-through rate (CTR) of the search results.
This experiment largely adheres to the Stable Unit Treatment Value Assumption (SUTVA), meaning that participants' outcomes were not significantly influenced by network interference or other unit's treatment~\citep{rubin1980randomization}. 
Furthermore, the likelihood of user-learning effects is minimal, given that the change is relatively implicit from the user's perspective and only one of many factors affecting recommendations is altered \citep{hohnhold2015focusing}.

\begin{figure}[h]
\centering
\includegraphics[width=0.4\textwidth]{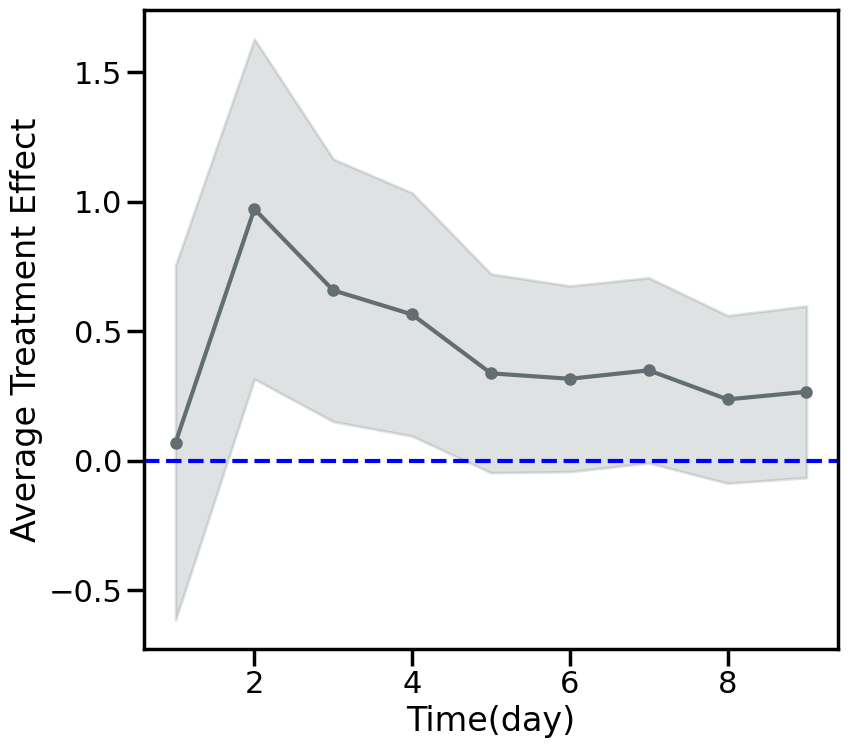}
\caption{Change of the average treatment effect over time.}
{\footnotesize \textit{Note}: 
The solid grey line represents the estimated effects with Difference-in-Means estimator from day 1 to day 9 for the empirical experiment, while the shaded region indicates the 95\% confidence interval. The horizontal dashed blue line at 0.0 denotes the null effect baseline.}
\label{fig:Empirical}
\end{figure}

This experiment lasted for 9 days involving 333,870 participants. 
As users interacted with the search engine being tested, they were randomly assigned to either the treatment or control group. The treatment group comprises 167,502 participants, while the control group includes 166,368 participants. We assess internal validity—the proper execution of the randomized experiment—using both a Sample Ratio Mismatch (SRM) test and an A/A test to ensure that the groups are comparable. Details are deferred to Web Appendix~\ref{sec:appendix:validity_check}.
Figure~\ref{fig:Empirical} displays the treatment effects observed among users up to the day they joined the platform. This suggests that the figure tracks how treatment effects evolve as the experiment’s duration increases and as more users joined. In this experiment, the treatment initially shows a significantly positive effect compared to the control condition; however, it is followed by a sharp decline, eventually stabilizing at a level that is statistically indifferent from zero. These findings indicate that experimenters might draw different conclusions depending on the experiment’s duration. Based on the first four days, one might conclude that the treatment is effective and should be launched, whereas the five to nine-day results suggest that the treatment has no significant effect, leading to the decision not to launch it.

\section{Identifying Sample Representativeness}
\label{sec:probSetup}

Our framework first aims to assess the sample's representativeness over time and to correct for any bias between the sample and the broader population. We begin by analyzing the sampling process throughout the experimental duration.

\subsection{Stages Identification}

Consider an online randomized experiment involving a population of \( N \) units (e.g., users). For each unit \( i \), researchers observe pre-treatment covariates \(\bm{X}_i \in \mathcal{X}\), which may include user demographics, historical behaviors, or other baseline characteristics. The experiment assigns units to one of two conditions: treatment (\( W_i = 1 \)) or control (\( W_i = 0 \)), where \( W_i \) denotes the binary treatment indicator. Let \( Y \) represent the outcome metric of interest, and \( Y_i(W_i = w) \) (or \( Y_i(w) \) for brevity) denote the potential outcome for unit \( i \) under treatment \( w \)\footnote{Our framework relies on the Stable Unit Treatment Value Assumption (SUTVA) \citep{cox1958planning, rubin1980randomization}, which we assume holds throughout the remainder of the paper.}.
The primary estimand of interest is the \textit{population average treatment effect} (PATE):
\begin{align}
    \tau = \mathbb{E}\left[Y_i(1) - Y_i(0)\right],
\end{align}



The above estimand cannot be directly calculated because, in practice, we only observe a sample of the population. Consequently, the causal effect must be inferred from a sample—a subset of the population that actually participated the experiment. In this paper, we focus exclusively on the nested trial design \citep{dahabreh2019extending}, which selects random samples from the target population.  

As we discussed, ongoing sampling makes it challenging to determine whether the current sample impartially represents the overall population throughout the experiment. For example, in a running experiment, users recruited in the initial days may disproportionately represent heavy users and may not accurately reflect WeChat’s full user base. Additionally, the treatment is likely to be more effective for active users, whose “previous-week” behaviors are more readily available. This uncertainty in sample composition can potentially lead to fluctuating estimations over time.

For the following discussion, let $t$ denote a specific time point within a discrete time horizon starting from zero to infinity, and $S_{it}$ denote an indicator determining whether unit $i$ is included in the experiment until time point $t$, with 1 indicating participation and 0 indicating non-participation.
The \textit{sample average treatment effect} (SATE) at each time point $t$ focuses on the estimate in the speciﬁc sample where units participate in: 
\begin{align}
\tau_t = \bbE[Y_i(1)-Y_i(0)| S_{it} = 1].
\end{align} 



The difference between $\tau$ and $\tau_t$ implies the bias in estimating the true effect on the target population using a limited sample, though it tends to diminish as more potential units become part of the experiment over time.
We can track the progress of the sampling process by observing $Pr[S_{it}=1]$, as a higher probability of inclusion for various units suggests a more representative sample. Theoretically, the bias is going to be eliminated when all units are included in the sample as time goes to infinity, which can be expressed as $\lim\limits_{t \to \infty} Pr[S_{it}=1] = 1$. 
Assuming that the outcome variable $Y$ is constant for a fixed unit $i$ and treatment $W_i=w$, the sample average treatment effect will eventually converge to an unbiased estimation of the target population average treatment effect:
\begin{align*}
\lim_{t \to \infty} \tau_t 
&= \lim_{t \to \infty} \bbE[Y_i(1)-Y_i(0)| S_{it} = 1] 
= \tau
\end{align*}

In practice, the experiment can only be conducted within a finite time horizon. Thus $\tau_t$ will always have a margin of error on a limited time scale. Nevertheless, it's possible to manage the estimation error within a constant bound.\footnote{The bias between $\tau_t$ and $\tau$ is contingent upon both the expected effect size among nonparticipants and the likelihood of participation in the experiment. In this context, we have made the inherent assumption that the former is bounded for all units, with our primary emphasis directed towards the latter consideration.} We define a specific tolerance level, denoted as $\rho$, whose value is positive and close to 0. As long as the bias between $\tau_t$ and $\tau$ shrinks to $\rho$, the sample is considered to be representative of the target population.



\begin{definition}[Time of Representativeness]
For some constant $\rho$, there exists a time point $T_r$, such that the absolute difference between the sample average treatment effect and the target population average treatment effect at the period after $T_r$ is smaller than $\rho$, i.e.,
$$|\tau_t - \tau|<\rho,\ \forall\ t>T_r$$
\end{definition}



The time point \(T_r\) indicates when the sample becomes representative---meaning that unbiased estimates from the sample can be applied to the target population. In other words, \(T_r\) marks the minimum duration after which the SATE approximately converges to the PATE. Notably, choosing a lower value for \(\rho\) imposes a more stringent threshold for the permissible bias between \(\tau_t\) and \(\tau\), and vice versa.
 

For experiments lasting less than \(T_r\), extrapolation-based methods can be employed to correct the sample selection bias that leads to unrepresentative samples.
The key idea behind these methods is to address discrepancies between the users participating in the experiment and the target population, which arise from distribution shifts in the sample covariates over time \citep{degtiar2023review}. Such methods typically require that the covariate distributions of the participating units do not deviate substantially from those of the target population \citep{imbens2015causal}.

Moreover, the probability of participating in the experiment varies with different pre-treatment covariates. In the early stages of the experiment, certain regions of the covariate space may be underrepresented---for instance, users with lower activity levels might join the experiment later. This issue tends to be mitigated as the experiment’s duration increases. Referring to the definition of strict overlap \citep{d2021overlap}, we identify a specific time point at which the overlap between the sample and the population in terms of covariate distributions is sufficiently high, thereby justifying the use of extrapolation-based methods to correct for selection bias.

\begin{definition}[Time of Overlap]
For some constant \(\eta_o\), there exists a time point \(T_o\) such that for all the time period after $T_o$, the probability of participating in the experiment given the pre-treatment covariates exceeds \(\eta_o\), i.e.
$$Pr[S_{it}=1 \vert \bm{X}_i = \bm{x}]>\eta_o,\ \forall\ t>T_o$$
\end{definition}\label{def:t_overlap}

Similar to the tolerance level \(\rho\) defined earlier, \(\eta_o\) is a parameter that regulates the level of overlap. The expression on the left-hand side, which represents the probability that units with given pre-treatment covariates are included in the experiment at or before time \(t\), is a crucial factor in developing the generalizing framework.  Let $\pi(t\vert \bm{X}_i) = Pr[S_{it}=1 \vert \bm{X}_i]$.
We will demonstrate in the following section that \(\pi(t \mid \bm{X}_i)\) serves as an important indicator that can be modeled using lifetime distribution functions from survival analysis \citep{klein2003survival}. With the aid of this model, \(T_o\) can be identified in conjunction with the parameter \(\eta_o\), thereby establishing a boundary for the validity of the adjusted estimation of the PATE.

\begin{figure}[h]
\centering
\includegraphics[width=0.8\textwidth]{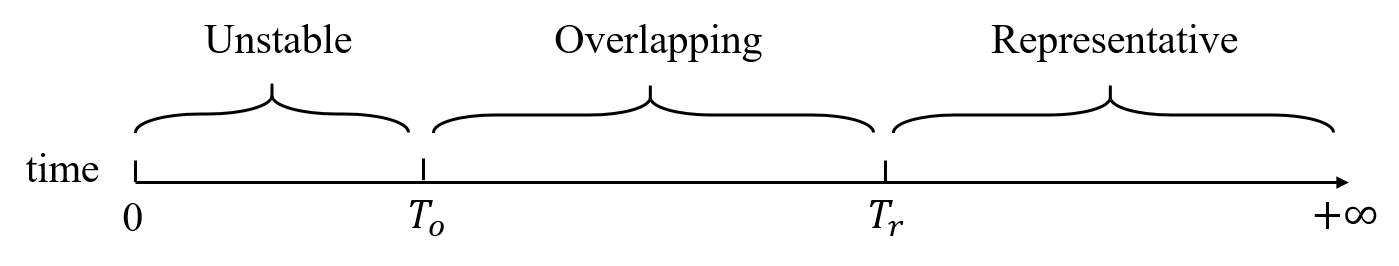}
\caption{Illustration of the different stages divided by criteria defined by specific time points.}
{\footnotesize \textit{Note}: The timeline of the experiment (along with the sampling process) is delineated into three stages: unstable stage, overlapping stage, and representative stage, by time points $T_o$ and $T_r$.}
\label{fig:TimePeriods}
\end{figure}

In summary, we find that the time horizon and the sampling process in the experiment are naturally divided into three stages by the two time points defined above. Before \(T_o\), no valid estimation of the PATE can be obtained from the sampled units, and continuing the experiments is recommended. Between \(T_o\) and \(T_r\), estimation can be achieved through appropriate extrapolation methods, although the precision and robustness of the estimates depend heavily on the chosen model. Beyond \(T_r\), the sampled units are representative of the target population, and the estimators for the SATE can be considered reliable and robust to the PATE.
See Figure~\ref{fig:TimePeriods} for an illustration of these three stages.

The time $T_r$ represents the ``representativeness'' extreme. It is the earliest point at which the remaining mismatch between the sample and the target population is provably bounded by $\rho$, such that simple estimators become valid for the target estimand without further adjustment. The estimators used at the representative stage typically exhibit lower variance and greater statistical power, as they require no reweighting—an advantage particularly relevant for operational decision-making.
More broadly, stopping at the overlapping stage enables earlier decisions using debiased estimators that account for sample representativeness, albeit at the cost of higher variance. In contrast, stopping at the representative stage sacrifices speed but yields more reliable inference for launch decisions. This trade-off allows different stakeholders to tailor stopping decisions according to their specific priorities.


\subsection{Identification Assumptions}

In this section, we discuss the three key assumptions required to identify the PATE at different stages of an experiment. The first assumption safeguards the internal validity of the experiment, while the other two ensure the generalizability of the experimental results.

\begin{assumption}[Strong ignorability]
For any unit $i$ that participated in the experiment at time period $t$, given the pre-treatment covariates, the treatment assignment is conditionally independent of the potential outcome, and the conditional probability of being assigned to either treatment should be positive, i.e.
\label{asp:ignorability}
\begin{align*}
(Y_i(1), Y_i(0))\independent W_i \vert \bm{X}_i, S_{it}=1\quad and\quad Pr[W_i=w \vert \bm{X}_i = \bm{x}, S_{it}=1]>0
\end{align*}
\end{assumption}

Assumption~\ref{asp:ignorability} is derived from the condition of strongly ignorable treatment assignment introduced by \cite{rosenbaum1983central}, which indicates the validity of causal inference drawn from the study's results. In the context of randomized experiments, this assumption is naturally satisfied since sampled units are completely randomly assigned to different treatments. 

\begin{assumption}[Conditional Exchangeability]\label{asp:exchangeability}
For any unit \(i\) at time \(t\), participation in the experiment is conditionally independent of the potential outcome given the pre-treatment covariates:
\[
(Y_i(1), Y_i(0)) \independent S_{it} \mid \bm{X}_i.
\]
\end{assumption}

Similar to the unconfoundedness condition in Rubin's causal model \citep{rubin1974estimating}, Assumption~\ref{asp:exchangeability} ensures that, for fixed covariate values, the distribution of the potential outcomes is consistent between the selected sample and the target population. Equivalently, 
non-participation (or missing data on outcomes for non-participants) is ``missing at random" \citep{little2019statistical}. The assumption implies that the probability of an outcome being missing depends only on the observed covariates and the treatment assignment does not affect the sampling process. This assumption partially ensures the possibility of inferring the PATE from sampled data using appropriate estimators.


\begin{assumption}[Positivity of Participation]\label{asp:positivity}
For every unit \(i\) and for all values of the pre-treatment covariates \(\bm{X}_i\), the probability of participating in the experiment at time \(t\) is strictly positive:
\[
\Pr[S_{it}=1 \mid \bm{X}_i = \bm{x}] > 0.
\]
\end{assumption}

Assumption~\ref{asp:positivity} guarantees that the participated units overlap all sets of conditions within observable covariates. This sufficient overlap allows us to reliably infer the treatment effect for a specific group, stratified by covariates, using the results from the same group within the selected samples. Notably, when \(t > T_o\), the positivity condition is automatically satisfied, ensuring that every subgroup defined by the observable covariates is represented in the sample. Assumption~\ref{asp:exchangeability} and Assumption~\ref{asp:positivity} collectively ensure that the experimental findings can be generalized to the target population.

\subsection{Estimation Methods}



We present two methods for estimating the PATE based on experimental results at different stages. It is important to note that for the period before \(T_o\), neither method is expected to yield reliable estimates. For the period after \(T_r\), the first method is typically employed, as it readily provides an unbiased estimation. Although the second method can also be applied after \(T_r\), it is generally reserved for the period between \(T_o\) and \(T_r\) to reduce potential errors resulting from the incomplete inclusion of covariate variables.

\subsubsection{Difference-in-Means Estimator}
The Difference-in-Means estimator, introduced by \citep{splawa1990application} as an unbiased estimator for the SATE, is widely used in practical applications. In our settings, for any time $t > T_r$, this estimator is deemed suitable for estimating the average treatment effect on the target population.  

\begin{align*}
\widehat{\tau}_{DIM}(t) = &\frac{1}{\sum_{i = 1}^N \bone\{S_{it} = 1,W_{i}=1\}}\sum_{i = 1}^N \bone\{S_{it}=1,W_{i}=1\}Y_i \\
 &- \frac{1}{\sum_{i = 1}^N \bone\{S_{it} = 1,W_{i}=0\}}\sum_{i = 1}^N \bone\{S_{it}=1,W_{i}=0\}Y_i
\end{align*}


\subsubsection{Inverse Probability Weighting (IPW) Estimator}

Derived from the Horvitz–Thompson estimator \citep{horvitz1952generalization}, the basic idea of a weighting-based estimator is to account for the distribution shift of outcomes between the sample and the target population by appropriately weighting the observations. This approach ensures that the weighted dataset is representative of the population. For instance, \cite{stuart2011use} introduced a propensity-score-based method to model the discrepancy between the sample and the target population, while \cite{dahabreh2019generalizing} proposed an inverse probability weighting (IPW) estimator that incorporates both the probability of participation and the treatment assignment. Drawing from these methods, we define the following IPW estimator tailored to the ongoing sampling process.


\begin{align*}
\widehat{\tau}_{IPW}(t) = &\frac{\sum_{i = 1}^N \bone\{S_{it} = 1\}}{N\cdot\sum_{i = 1}^N \bone\{S_{it} = 1,W_{i}=1\}}\sum_{i = 1}^N \frac{\bone\{S_{it}=1,W_{i}=1\}}{\hat{\pi}(t\vert\bm{X}_i)}Y_i \\
 &- \frac{\sum_{i = 1}^N \bone\{S_{it} = 1\}}{N\cdot\sum_{i = 1}^N \bone\{S_{it} = 1,W_{i}=0\}}\sum_{i = 1}^N \frac{\bone\{S_{it}=1,W_{i}=0\}}{\hat{\pi}(t\vert\bm{X}_i)}Y_i.
\end{align*}

where $\hat{\pi}(t\vert\bm{X}_i)$ is an estimator for $\pi(t\vert\bm{X}_i)$. 

\noindent
The proof of the unbiasedness of this estimator is presented in Appendix~\ref{sec:appendix:unbiasProof}. Apart from the reweighting method, alternative approaches—such as the outcome regression method and the doubly robust method—can also be employed to generate adjusted estimators \citep{ding2018causal}. Compared to these methods, the IPW estimator is model-free, requiring no assumptions about the form of the outcome variable, making it more adaptable to a wide range of scenarios. Although the outcome regression method and the doubly robust method are less frequently used in practice, we provide an overview of them and primarily focus on IPW estimation (see Appendix~\ref{sec:appendix:estimators}).

\subsubsection{Inference}
Inference for the Difference-in-Means estimator is typically performed using a two-sample t-test. In contrast, for the IPW estimator, the variance can be estimated using the sandwich estimator, which leverages meta-level statistics \citep{freedman2006so}. However, to avoid the potentially conservative variance estimates that may arise with the sandwich estimator, the bootstrap approach is often preferred in practice \citep{efron1994introduction}. In our empirical studies, we demonstrate the use of the bootstrap method for inference. A potential concern with the bootstrap for IPW estimators is the ``generated regressor'' problem. Because the participation probabilities $\hat{\pi}(t \mid \bm{X}_i)$ are themselves estimated, standard nonparametric bootstrap may understate the true variability if it does not account for the estimation uncertainty in the weights \citep{abadie2008failure}. To mitigate this, we re-estimate the survival model within each bootstrap replicate, so that both the weights and the treatment effect estimate are recomputed jointly. This approach captures the full sampling variability of the two-step procedure and has been shown to yield valid inference in similar IPW settings \citep{hirano2003efficient}.



\section{Heuristic Method}
\label{sec:heuristics}

Thus far, we have described the three stages of the ongoing sampling process in randomized experiments and the corresponding estimation strategies for each period. In the following section, we first introduce the specific survival analysis function used to model \(\pi(t \mid \bm{X}_i)\), from which the IPW estimator is naturally derived. We then demonstrate how to use \(\hat{\pi}(t \mid \bm{X}_i)\) to empirically determine the time points \(T_o\) and \(T_r\).

\subsection{Survival Analysis}

Survival analysis studies the time until an event occurs and how that time varies with observed conditions \citep{jenkins2005survival}. In biomedical applications, the event is often death; in our setting, the event is \emph{participation}, meaning that a user is first exposed to the experiment. Accordingly, we model each user’s \emph{time-to-participation}. At any calendar time \(t\), users who have already participated contribute observed outcomes, whereas users who have not yet been exposed remain ``at risk'' with unobserved exposure times. This feature of online experimentation induces natural right-censoring: if a user has not participated by time \(t\), then their participation time exceeds \(t\) but is otherwise unknown. Survival analysis is designed for exactly this type of time-to-event data with censoring, making it well suited for modeling the ongoing selection process in online experiments.

In this study, survival analysis is preferred over conventional static approaches for estimating participation probabilities. A natural alternative would be to fit a cross-sectional binary classifier (e.g., logistic regression) at each time point to predict whether a user has participated. However, such approaches discard the temporal ordering of participation events, treat right-censored observations as simple non-events, and require refitting at each monitoring point without leveraging the sequential structure of the data. In contrast, survival models natively incorporate censoring into the likelihood, exploit the full trajectory of participation timing, and produce monotonically increasing participation probability estimates---consistent with the irreversible nature of experiment entry. These properties make survival analysis not merely a convenient modeling choice but the methodologically appropriate framework for the ongoing sampling problem we address.

Formally, let \(T\) denote the time until participation. We can define the survival function \(S(t \mid \bm{X})\) and the lifetime distribution function \(F(t \mid \bm{X})\) as follows.\footnote{To avoid confusion with the participation indicator \(S_{it}\) defined in Section~\ref{sec:probSetup}, we note that \(S(t \mid \bm{X})\) denotes the survival function from survival analysis, following standard notation in that literature. The two are related: \(S_{it} = \mathbf{1}\{T_i < t\}\), so \(\Pr[S_{it}=1 \mid \bm{X}_i] = F(t \mid \bm{X}_i) = 1 - S(t \mid \bm{X}_i)\).}
\[
S(t \mid \bm{X}) = \Pr(T \geq t \mid \bm{X}), \qquad
F(t \mid \bm{X}) = 1 - S(t \mid \bm{X}) = \Pr(T < t \mid \bm{X}).
\]

Consider \(T_i\) as the value of \(T\) for a specific unit \(i\). Notice that the indicator \(S_{it}\) can be written as $S_{it} = \mathbf{1}\{T_i < t\}.$
Thus, we deduce that \(\pi(t \mid \bm{X}_i)\) is simply the realization of the lifetime distribution function for a unit with covariate profile \(\bm{X}_i\):\[
\pi(t \mid \bm{X}_i) = \Pr\big(\mathbf{1}\{T_i < t\} = 1 \mid \bm{X}_i\big) = F(t \mid \bm{X}_i).
\]

This equivalence links external-validity adjustment under sequential exposure to estimating a time-to-event distribution. We thus employ survival-analysis methods to estimate \(\pi(t \mid \bm{X}_i)\). In what follows, we introduce a nonparametric approach based on the Kaplan--Meier estimator\footnote{Our framework accommodates a broad class of survival models, including the Cox proportional hazards model; see Web Appendix~\ref{sec:appendix:cox}.}.


\subsubsection{Kaplan-Meier Estimator}

The Kaplan–Meier estimator is a non-parametric method used to estimate the survival function. The resulting Kaplan–Meier survival model is a step function that decreases monotonically over time. Let 
$
0 \leq t_1 < t_2 < \cdots < t_j < \cdots
$
denote the discretized time points. Given a covariate profile \(\bm{X}_i\), we define the Kaplan–Meier estimator of the survival function as follows:

\begin{align*}
\hat{S}(t|\bm{X}_i = \bm{x}) = \prod_{j: t_j\in [0,t]} 
\left(1 - \frac{\sum_{i = 1}^N\bone\{S_{it_j}=1,\bm{X}_i = \bm{x}\} - \sum_{i = 1}^N\bone\{S_{it_{j-1}}=1,\bm{X}_i = \bm{x}\}}{N - \sum_{i = 1}^N\bone\{S_{it_{j-1}}=1,\bm{X}_i = \bm{x}\}}\right).
\end{align*}

With the Kaplan-Meier estimator, we can obtain $\hat{\pi}(t\vert\bm{X}_i)$ with the plug-in approach:
\begin{align*}
& \hat{\pi}_{KM}(t\vert\bm{X}_i= \bm{x}) = \hat{F}(t\vert\bm{X}_i= \bm{x}) \\
= & 1 - \prod_{j: t_j\in [0,t]} 
\left(1 - \frac{\sum_{i = 1}^N\bone\{S_{it_j}=1,\bm{X}_i = \bm{x}\} - \sum_{i = 1}^N\bone\{S_{it_{j-1}}=1,\bm{X}_i = \bm{x}\}}{N - \sum_{i = 1}^N\bone\{S_{it_{j-1}}=1,\bm{X}_i = \bm{x}\}}\right).
\end{align*}

\subsection{Determination Strategies}
\label{sec:determineStrategy}


In the subsequent sections, we will illustrate the relationship between \(\hat{\pi}(t \mid \bm{X}_i)\) and the criteria used to differentiate the sampling stages, and we will provide heuristic strategies for determining \(T_o\) and \(T_r\) in practice.

\subsubsection{Time of Overlap}

Given the definition, \(T_o\) should be the time point where \(\hat{\pi}(t \mid \bm{X}_i=\bm{x}) > \eta_o\) for all \(t > T_o\) and for every possible covariate vector in \(\bbX\). Alternatively, we define
\[
\hat{\pi}^{inf}(t) = \inf_{\bm{x} \in \bbX} \hat{\pi}(t \mid \bm{X}_i=\bm{x}) > \eta_o\ 
\]
which transforms the determination of \(T_o\) into the task of finding the smallest \(t\) such that \(\hat{\pi}^{inf}(t) > \eta_o\). This ensures that even the covariate group with the lowest participation probability is adequately covered. We provide further discussion on the chosen of \(\eta_o\) in practice in Section~\ref{sec:parameterSelect}.

We consider \(\hat{\pi}^{inf}(t)\) as a heuristic function for distinguishing the sampling stages. To streamline the process, we compute \(\hat{\pi}^{inf}(t)\) at each time \(t\) rather than aggregating the estimated participation probabilities across different values of \(\bm{X}\). We then continue to use \(\hat{\pi}^{inf}(t)\) to guide the determination of \(T_r\) through its definition.

\subsubsection{Time of Representativeness}

When considering the periods beyond $T_o$, since $\widehat{\tau}_{DIM}(t)$ and $\widehat{\tau}_{IPW}(t)$ are unbiased estimators for $\tau_t$ and $\tau$ respectively, the upper bound of the absolute difference between these two estimands can be directly calculated. We revisit the definition of time of representativeness and  develop the following proposition.

\begin{proposition}[Upper Bound of Bias]
\label{prop:UpperBound}
Consider completely randomized experiments across heterogeneous groups characterized by $\bm{X}$, for all $t>T_o$, the estimated absolute difference between $\tau_t$ and $\tau$ is less than the product of $\hat{\pi}_{inf}(t)$ and the weighted average of the absolute values of heterogeneous treatment effect, i.e.
\begin{align*}
|\widehat{\tau}_{DIM}(t) - \widehat{\tau}_{IPW}(t)|
\leq 2\cdot\left(\frac{1}{\hat{\pi}^{inf}(t)}-1\right)\cdot\sum_{\bm{x}\in\bbX}\frac{\sum_{i=1}^N\bone\{S_{it_j}=1,\bm{X}_i = \bm{x}\}}{\sum_{i=1}^N\bone\{S_{it_j}=1\}}|\hat{\tau}_{HTE}(t,\bm{x})|,\ \forall\ t>T_o
\end{align*}
where
\begin{multline*}
\hat{\tau}_{HTE}(t,\bm{x}) = \frac{1}{\sum_{i = 1}^N \bone\{S_{it} = 1,\bm{X}_i = \bm{x},W_{i}=1\}}\sum_{i = 1}^N \bone\{S_{it}=1,\bm{X}_i = \bm{x},W_{i}=1\}Y_i \\
 - \frac{1}{\sum_{i = 1}^N \bone\{S_{it} = 1,\bm{X}_i = \bm{x},W_{i}=0\}}\sum_{i = 1}^N \bone\{S_{it}=1,\bm{X}_i = \bm{x},W_{i}=0\}Y_i.
\end{multline*}
Moreover, if we assume that the weighted average of the sum of $|\hat{\tau}_{HTE}(t,\bm{x})|$ is bounded by $\frac{C}{2}\cdot |\widehat{\tau}_{DIM}(t)|$ where $C$ is a positive constant, then $T_r$ can be identified as the smallest $t$ that satisfies
$$\hat{\pi}^{inf}(t)>\frac{C\cdot |\widehat{\tau}_{DIM}(t)|}{\rho + C\cdot |\widehat{\tau}_{DIM}(t)|}$$

\end{proposition}

The proof of Proposition~\ref{prop:UpperBound} is provided in Appendix~\ref{sec:appendix:proof1}. Based on Proposition~\ref{prop:UpperBound}, we are able to associate $T_r$ with the heuristic function $\hat{\pi}_{inf}(t)$ and consequently determine it. Let 
\begin{align}
\eta_r = \frac{C\cdot \sup\limits_{t}\left\{|\widehat{\tau}_{DIM}(t)|\right\}}{\rho + C\cdot \sup\limits_{t}\left\{|\widehat{\tau}_{DIM}(t)|\right\}},
\label{eqn:thresholdRepresent}
\end{align}
the problem of determining $T_r$ is reframed as finding the smallest $t$ for which $\hat{\pi}^{inf}(t)$ exceeds a certain constant $\eta_r$.

It is important to note the gap between the theoretical Definition~1, which characterizes $T_r$ in terms of the unobservable true quantities $\tau_t$ and $\tau$, and its empirical operationalization through $\hat{\pi}^{inf}(t)$. The heuristic function provides an upper bound on the bias $|\tau_t - \tau|$ via Proposition~\ref{prop:UpperBound}, rather than a direct estimate of the bias itself. This means the empirically identified $T_r$ is conservative: it guarantees that the bias is at most $\rho$, but the actual bias may be substantially smaller. The quality of this approximation depends on (i)~how tightly the constant $C$ bounds the ratio of heterogeneous to average effects, and (ii)~the accuracy of the survival model's estimates of participation probabilities. 

\subsection{Selection of Thresholds}
\label{sec:parameterSelect}

We identify that, given two pre-determined thresholds \(\eta_o\) and \(\eta_r\), the stage boundaries \(T_o\) and \(T_r\) can be derived by comparing the value of \(\hat{\pi}^{\text{inf}}(t)\) with these thresholds. In this section, we discuss how to determine \(\eta_o\) and \(\eta_r\).

\subsubsection{Selection of \(\eta_o\)}

The value of \(\eta_o\) governs the overlap quality between the sample and the target population considering the covariate variables. We can observe that the strictest approach on determining \(\eta_o\) is to set \(\eta_o = 0.5\), which bounds the probability of participation given covariate group \(\bm{X}\) far away from zero. Intuitively, it means that for every group characterized by \(\bm{X}\), the probability of participation exceeds the probability of nonparticipation. 

Increasing \(\eta_o\) strengthens the overlap guarantee and generally leads to more robust estimators, but it also implies a longer waiting time before the overlap condition is met. In practice, it is advisable to choose \(\eta_o\) close to \(0.5\) to reduce the risk of misidentifying the overlap stage, which could potentially inflate the Type I error rate of the effect estimation. Nonetheless, our method is to some extent robust to the choice of \(\eta_o\). Table~\ref{tab:sensitivityCheck} in the Web Appendix reports the bias and MSE of the IPW estimator at time \(T_o\) for different values of \(\eta_o\) under the simulation setting introduced in Section~\ref{sec:simulation}, and shows that both bias and MSE change only modestly when \(\eta_o\) is set slightly below \(0.5\).

\subsubsection{Selection of \(\eta_r\)}

Intuitively, since \(\widehat{\tau}_{DIM}(t)\) is equivalent to the weighted average of the heterogeneous treatment effects \(\hat{\tau}_{HTE}(t,\bm{x})\), the constant \(C\) will exceed two only if there exist heterogeneous groups with treatment effects in the opposite direction to the overall sample average treatment effect. This phenomenon often occurs when a treatment has an extremely pronounced negative impact on a minority group, warranting extra caution. For example, treatments tailored to appeal to a specific subset of users may be attractive to that group but may irritate the majority. Such treatments may show negligible or even negative effects on the overall population, while experiments with biased samples might fail to detect this. We have observed that the multiplier \(C\) typically is around two as the sample approaches representativeness (see Appendix~\ref{sec:appendix:C}). In practice, \(C\) is often predetermined based on prior knowledge from historical experiments.

Determining $\eta_r$ empirically involves a tradeoff similar to that encountered in determining $\eta_o$, specifically balancing experiment time against the credibility of the estimator. By eliminating the influence of the magnitude of the treatment effect across different experiments, we can express $\eta_r$ as:
$$\eta_r = \frac{C}{\frac{\rho}{\sup\limits_{t}\left\{|\widehat{\tau}_{DIM}(t)|\right\}} + C}.$$ 
Considering the allowable bias $\rho$ as a proportion of $\sup\limits_{t}\left\{|\widehat{\tau}_{DIM}(t)|\right\}$, we observe that $\eta_r$ is a probability value less than one. It is larger given a stricter tolerance level (smaller $\rho$), indicating that a longer duration of the experiment is required to achieve a more rigorous representative sample, and vice versa. Alternatively, $\eta_r$ can be interpreted as a parameter to control the possible bias for experiments terminated at time after $T_r$. Based on the above equation, the absolute difference between $\tau_t$ and $\tau$ is bounded by:


\begin{align*}
|\tau_t - \tau| < \rho  = \frac{1-\eta_r}{\eta_r} \cdot C \cdot \sup\limits_{t}\left\{|\widehat{\tau}_{DIM}(t)|\right\} 
\end{align*}

Once $\eta_r$ is determined, one would
expect the observed difference-in-means to change by no more than $\frac{1-\eta_r}{\eta_r} \cdot C$ of the absolute value of the observed difference-in-means at time after $T_r$.

Together, the choices of \(\eta_o\) and \(\eta_r\) govern a bias-variance tradeoff that is central to the practical deployment of our framework. Higher thresholds reduce estimation bias by enforcing stronger overlap or representativeness, but increase both the required experiment duration and the variance of the estimator at the time the threshold is first met, since fewer observations accumulate by that point in some covariate strata. Conversely, lower thresholds allow earlier stopping but accept greater residual bias. A detailed sensitivity analysis examining how these thresholds affect bias and MSE under both the Kaplan-Meier and Cox models is provided in Web Appendix~\ref{sec:appendix:sensitivity_analysis}.

\section{Experiments and Results}
\label{sec:results}




In this section, we evaluate our proposed framework at two levels. We begin with a running example---a real-world online experiment conducted on WeChat. We then demonstrate platform-scale applicability by applying the framework to 600 WeChat A/B tests. Finally, we quantify its operational impact by comparing experimenters' stopping behavior before and after the method was deployed on the platform. To further illustrate the estimation details, We  validate the approach in a controlled synthetic setting in Appendix~\ref{sec:simulation}.




\subsection{A Real-World Experiment}
Before formally applying our methodology, we illustrate the sample distribution shift phenomenon in the running example by dividing the population into eight subgroups based on a key feature, labeled from level 0 to level 7.\footnote{According to the non-disclosure agreement, we are not able to reveal physical meanings about the feature.} Each subgroup represents users with a specific level of this feature. We observe that higher subgroup levels correspond to users with heavier content consumption. Intuitively, the performance of the recommendation algorithm tested in this experiment is likely influenced by users' content consumption levels, as these levels directly affect the accumulation of historical behavior records used to generate accurate recommendations.



\begin{figure}[h]
\centering
\includegraphics[width=0.45\textwidth]{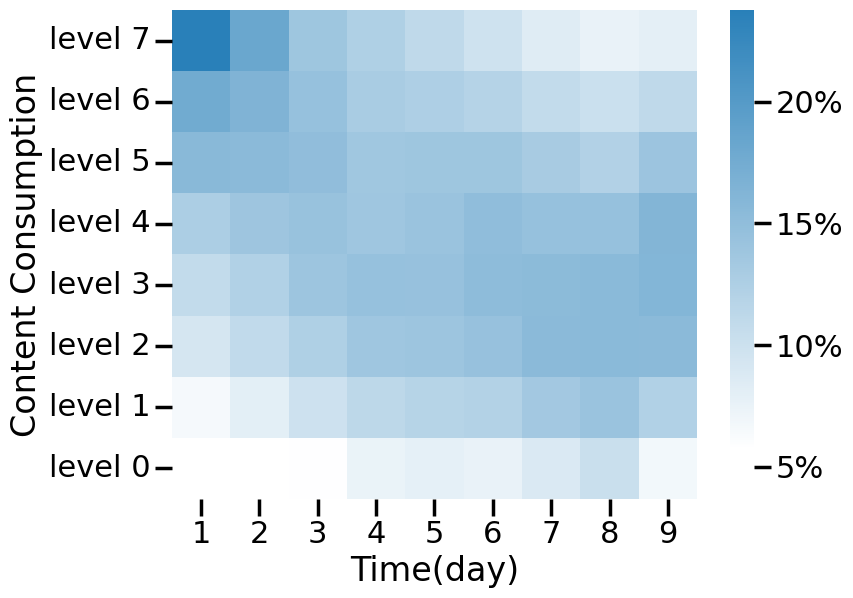}
\caption{Changes in the covariate distribution within the sample over the course of 9 days.}
{\footnotesize \textit{Note}: 
The x-axis denotes the time in days, while the y-axis categorizes the selected sample in experiment into 8 subgroups based on the content consumption levels from 0 to 7. The color intensity, ranging from light blue to dark blue, indicates the proportion of the subgroup users in sample, with darker shades representing higher percentages.}
\label{fig:sample_dist}
\end{figure}

Figure~\ref{fig:sample_dist} presents a heat map that illustrates changes in the subgroup distribution over the experimental period. We observe that the proportion of users with high content consumption (levels 5–7) gradually decreases over time, while the proportion of dormant users increases. Given that heavy and light users may experience different treatment effects, the overall sample average treatment effect can fluctuate over time. In this experiment, the treatment involves an algorithm that ranks search results by prioritizing content consumed in the previous week. This treatment tends to affect heavy users more, as their richer historical behavior provides more data for the algorithm. Combined with the overrepresentation of heavy users in the early stages of the experiment, this partially explains why the treatment effect is particularly large during the first few days (see Figure~\ref{fig:Empirical}).
Overall, this example highlights that external validity issues caused by the ongoing sampling process of online experiments can arise from a combination of shifting covariate distributions over time and heterogeneous effects across subgroups.

Next, we apply our phased debiasing framework throughout the experimental period to adjust the estimation. We begin by identifying key covariates to characterize experimental units (i.e., users). Based on domain knowledge, we select variables associated with both user participation status and click-through rate, such as average login days per week, average query frequency per day, and user demographics (e.g., gender, age, and education level), as covariates in our framework. Experimenters set thresholds based on their business knowledge, tolerance for bias and preliminary calculations. \(\eta_o\) is set to \(0.5\), indicating a higher probability of participation against nonparticipation after the overlapping stage; \(\eta_r\) is set to \(0.9\), which is determined according to equation~\eqref{eqn:thresholdRepresent}, where the constant \(C = 2\) precalculated from historical
experiments, and \(\rho = 0.2\cdot\sup\limits_{t\to\infty}\left\{|\widehat{\tau}_{DIM}(t)|\right\}\) reflects a tolerance for a 20\% relative estimation error in the experiments. (see Section~\ref{sec:parameterSelect} for details). 

The heuristic function \(\hat{\pi}^{inf}(t)\) is derived from a Kaplan–Meier model based on the covariates discussed in the previous sections.
At each time point \(t\), \(\hat{\pi}^{inf}(t)\) is calculated and compared against the thresholds \(\eta_o\) and \(\eta_r\) to determine the current stage of the experiment. During the overlapping stage, the IPW estimator is used to adjust for bias, while the Difference-in-Means estimator continues to be applied during the representative stage.

\begin{figure}[h]
\centering
\includegraphics[width=0.4\textwidth]{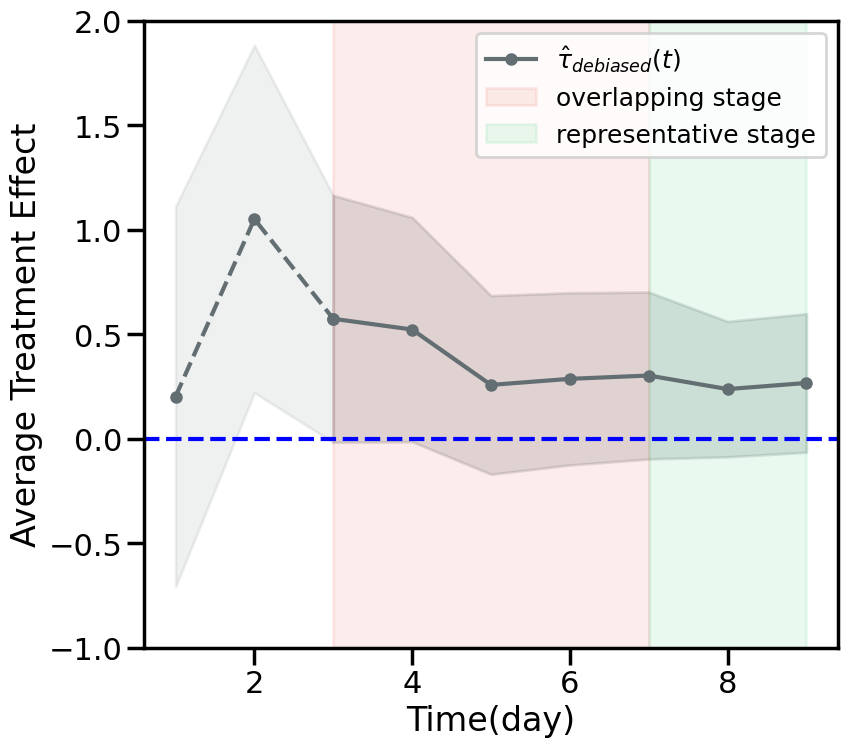}
\caption{Debiased estimation of the average treatment effect at different experiment stages.}
{\footnotesize \textit{Note}: 
The solid grey line represents the debiased estimated effects, with shadows indicating 95\% confidence intervals. The red and green shaded regions correspond to the overlapping and representative stages of the experiment, respectively. The horizontal dashed blue line at 0.0 represents the null effect baseline.}
\label{fig:sample_debiased}
\end{figure}

Figure~\ref{fig:sample_debiased} presents the debiased estimation of the ATE across different experimental stages. The period prior to the overlapping stage yields unreliable estimates, as illustrated in Section~\ref{sec:heuristics}; therefore, estimates from this phase should not be used to infer the PATE or guide product decision-making. In other words, experimenters should refrain from stopping experiments during the unstable stage unless there are compelling operational reasons.

The debiasing is applied only to the latter two stages: the overlapping stage and the representative stage. The figure shows that our method enables the experiment to reveal the final, stabilized outcome (i.e., no significant treatment effect) as early as the beginning of the overlapping stage—two days earlier than the unadjusted estimator (as shown in Figure~\ref{fig:Empirical}). This suggests that product decisions can be made reliably based on experiments concluded in either the overlapping or the representative stage.

Moreover, Figure~\ref{fig:sample_debiased} shows a clear reduction in estimator variance from the overlapping stage to the representative stage, while the bias in the later part of the overlapping stage is comparable to that in the representative stage. Due to its weighting structure, the IPW estimator exhibits higher variance than the difference-in-means estimator. The latter is therefore better aligned with the decision-making objective—namely, achieving higher statistical power, enabling more reliable inference, and reducing the risk of overlooking potentially effective treatments. 
This result further underscores the importance of reaching the representative stage, where simpler estimators can be applied reliably with improved statistical efficiency.

\begin{figure}[h]
\centering
\includegraphics[width=0.45\textwidth]{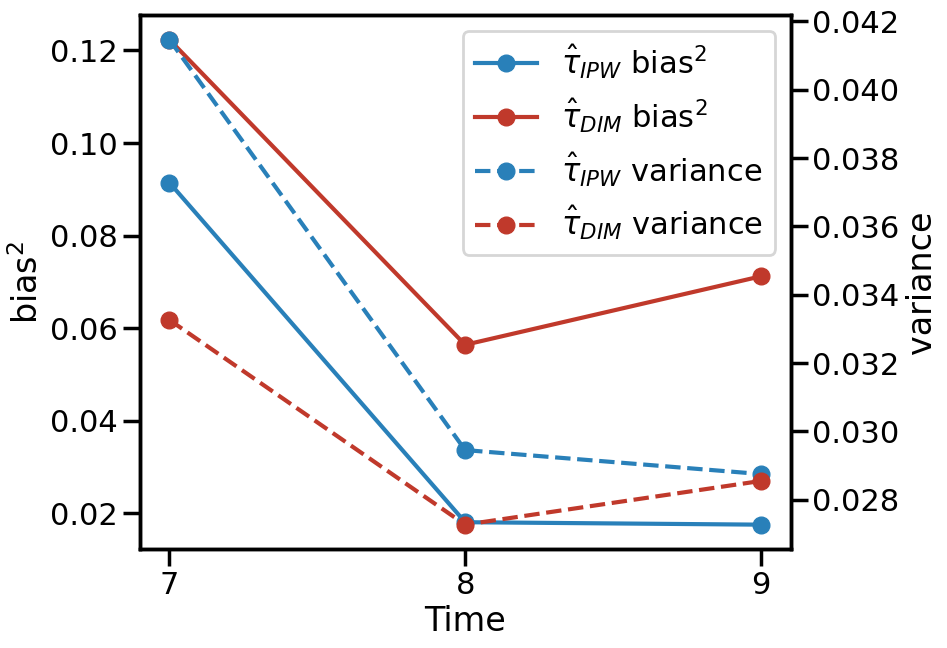}
\caption{Bias-variance trade-off for two estimators during the representative stage over time.}
\label{fig:bias_var}
\end{figure}

It is important to note that while both \(\hat{\tau}_{IPW}(t)\) and \(\hat{\tau}_{DIM}(t)\) can be used to estimate the average treatment effect during the representative stage, they differ in their finite-sample performance, giving rise to a bias--variance trade-off in practice. Figure~\ref{fig:bias_var} illustrates the evolution of squared bias and variance for these two estimators during the representative stage (day 7 to day 9). The results show that \(\hat{\tau}_{IPW}(t)\) exhibits lower bias but higher variance compared to \(\hat{\tau}_{DIM}(t)\).
Given the critical importance of variance reduction in practical industry settings, we recommend \(\hat{\tau}_{DIM}(t)\) as the preferred estimator for the average treatment effect in the representative stage. 

\subsection{Platform Application}



To assess how our methods scale in platform operations, we first applied them to 600 experiments randomly selected from the ``overdone" experiments conducted across various businesses on WeChat. The selection of these experiments was based on the following criteria:
1. These experiments satisfy SUTVA, the basic assumption underlying our method.
2. The average treatment effects stabilized as the experiment duration was extended, with an average duration of over three weeks.
These experiments, which happened to run for extended durations, allow us to reasonably assume that the stabilized effects observed at the end represent the true treatment effect (i.e., the ``ground truth" for PATE). This offers an opportunity to evaluate our method's ability to mitigate sampling bias and identify more effective treatments without requiring lengthy trials.
\footnote{Note that these experiments were not extended in duration because of our study; instead, WeChat conducts thousands of experiments per day, and we collected data from those that happened to be overdone for various practical reasons.}\footnote{We acknowledge that the 600 experiments used in this empirical analysis are subject to sample selection and lack full representativeness. This limitation is inherent in our setting, as obtaining a reliable ground truth for benchmarking requires working with such selected samples.}\footnote{Our approach is based on the logic that if the bias due to sample representativeness is corrected, the estimated effects should be closer to the ``ground truth''. However, even if sample representativeness is improved, other factors may still contribute to the gap between the estimated effects and the “ground truth” — the stabilized effects. Therefore, the effectiveness measured by the following metrics should be considered a lower bound.}

We evaluate the effectiveness of our method using two performance metrics: the False Positive Rate (FPR) and the True Positive Rate (TPR). 
Similar to the confusion matrix used in machine learning \citep{stehman1997selecting}, we define four possible outcomes for a statistical test conducted when the experiment is terminated at a specific period. A \textit{True Positive} (TP) occurs when a test correctly indicates that an effective treatment is significantly positive, whereas a \textit{True Negative} (TN) occurs when a test correctly indicates that an ineffective treatment has no significant effect. Conversely, a \textit{False Positive} (FP) occurs when a test incorrectly indicates that an ineffective treatment is significantly positive, and a \textit{False Negative} (FN) occurs when a test incorrectly indicates that an effective treatment has no significant effect. The False Positive Rate is calculated as 
$
FPR = \frac{FP}{FP + TN},
$
which represents the probability that a non-effective treatment is mistakenly identified as having a positive effect (and potentially launched as a product). The True Positive Rate is calculated as 
$
TPR = \frac{TP}{TP + FN},
$
which represents the probability that an effective treatment is successfully detected. Our objective is to assess whether our method can increase the TPR while not increasing the FPR, thereby enhancing the overall efficacy of decision-making based on the experimental results. Based on the estimated effects at the conclusion of the experiments --- the ``ground truth'', we discover that out of the 600 experiments conducted, 510 reported insignificant effects, while 90 yielded significant results at a significance level of $\alpha = 0.05$. 


\begin{figure}[h]
\centering
\includegraphics[width=0.9\textwidth]{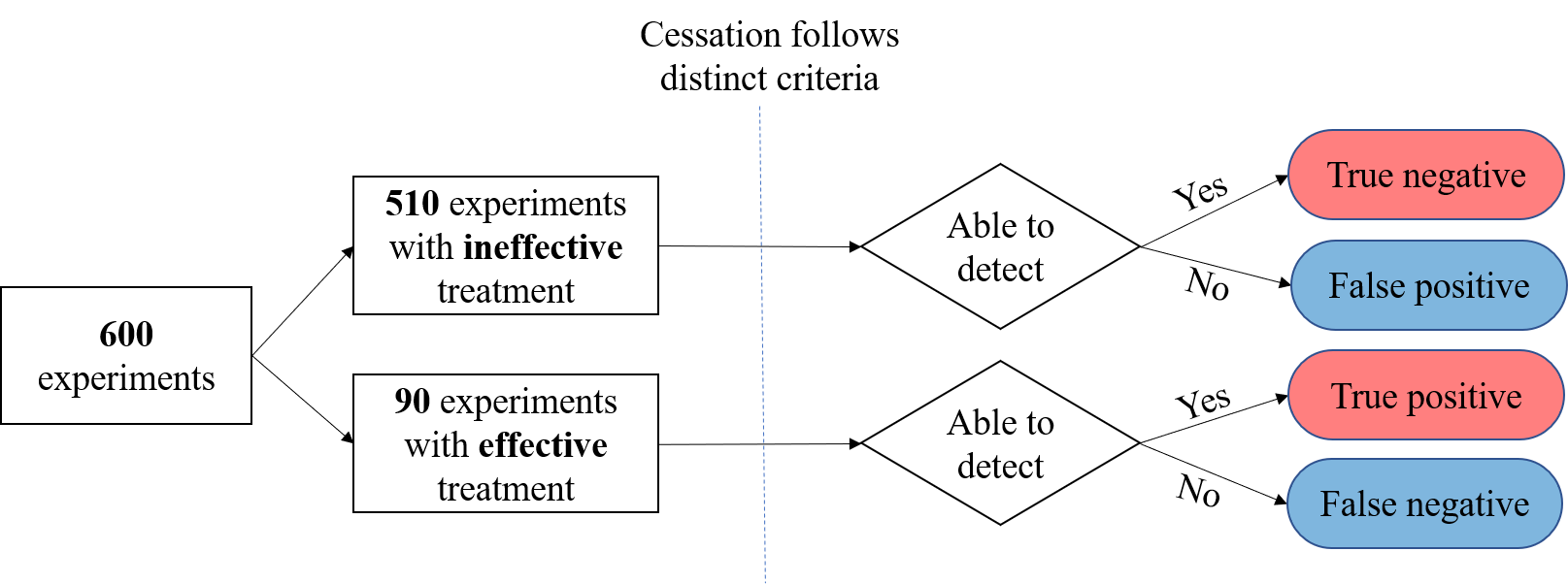}
\caption{Illustration of categorization of empirical experiments.}
\label{fig:flowchart_empirical}
\end{figure}

We compare our framework with the commonly used baseline method for determing sample size and experiment duration based on power analysis~\citep{deng2021post,xiang2022multi}. Although this approach is widely adopted in practice, it does not account for external validity; instead, it guarantees that the sample size is large enough to achieve the desired power for a two-sample \emph{t}-test. Specifically, at a 95\% confidence level and 80\% power, the minimum required sample size is given by
$
n = \frac{16\sigma^2}{\Delta^2},
$
where \(\sigma^2\) represents the variance of the outcome of interest, and \(\Delta\) denotes the smallest treatment effect the experimenters aim to detect \citep{kohavi2009controlled}.

Table~\ref{tb:example2} presents a performance comparison between our method and the baseline approach, where experiment duration is determined solely by power analysis. The first column displays the results of the baseline approach, while the other three columns summarize our method's outcomes in terms of FPR and TPR, as if the experiments had stopped at different stages.
More specifically, the second column illustrates the scenario where the experiment is concluded at the \emph{overlapping stage}, determined by the threshold $\eta_o = 0.5$, with the IPW estimator employed for inference. The third and fourth columns show the experiment being terminated at the \emph{representative stage}, determined by thresholds $\eta_r = 0.8$ and $\eta_r = 0.9$ respectively\footnote{The thresholds here are calculated based on equation~\eqref{eqn:thresholdRepresent}, with the constant \(C = 2\) precomputed for all experiments in this setting and \(\rho = \{0.2,0.5\}\cdot\sup\limits_{t\to\infty}\left\{|\widehat{\tau}_{DIM}(t)|\right\} \).}, using the Difference-in-Means estimator for statistical testing. Our framework does not recommend concluding experiments in the unstable stage.

We observe that our method consistently outperforms the baseline approach when stopping at either the overlapping or representative stages. Overall, our method increases the TPR by approximately $28$–$37\%$ while reducing the FPR by about $17$–$29\%$. This substantial improvement demonstrates that our method can identify more effective treatments without misclassifying a greater number of ineffective ones.

\begin{table}[h]
\centering
\caption{Performance of Different Methods in Terms of FPR and TPR}
\label{tb:example2}
\begin{tabular}{c|cccc}
\toprule
 & \textbf{Baseline} & \textbf{Overlapping} & \multicolumn{2}{c}{\textbf{Representative}} \\
 & \textbf{Approach} & $(\eta_o = 0.5)$ & $(\eta_r = 0.8)$ & $(\eta_r = 0.9)$ \\
\midrule
\textbf{False Positive Rate (FPR)} & 11.3\% & 9.4\% & 8.8\% & 8.0\% \\
\textbf{True Positive Rate (TPR)}  & 35.6\% & 45.5\% & 48.9\% & 48.9\% \\
\bottomrule
\end{tabular}
\end{table}


Furthermore, we investigate the sources of these improvements. By stratifying experiments based on the stage they have reached at the time they first satisfy the baseline stopping criterion, we find that more than 95\% reach the overlapping stage (with \(\eta_o = 0.5\)), and more than 70\% reach the representative stage. These proportions remain essentially unchanged when comparing \(\eta_r = 0.8\) to \(\eta_r = 0.9\), supporting the robustness of our chosen thresholds.
This decomposition suggests that the gains in TPR and FPR primarily arise through two channels: (i) avoiding estimates generated during the unstable stage (5\% of experiments), and (ii) improving inference for experiments that stop in the overlapping stage (25\% of experiments). In total, the framework materially enhances treatment effect estimation and operational decision-making for approximately 30\% of experiments when they are concluded at the representative stage.
Beyond improving the detection of effective treatments, it also provides experimenters with actionable, real-time diagnostics of sample representativeness throughout the experimental process.

Additionally, our method has been integrated as a function within the A/B testing system at WeChat in 2025. This platform-wide deployment enables us to demonstrate the practical impact of our approach on experimenters’ stopping decisions and experiment duration. We compare experimentation outcomes over a 12-month period before deployment with those over a 12-month period after deployment\footnote{Due to the nondisclosure agreement (NDA), we are unable to disclose the analysis details, as doing so may inadvertently reveal sensitive platform information prohibited under the NDA.}.
We find that the average experiment duration increases from 8.21 days before deployment to 9.91 days after deployment, representing a 20.7\% increase. This pattern suggests that experimenters actively account for potential external validity concerns and are willing to trade off iteration speed for improved sample representativeness. 

We further observe a significant shift in the stages at which experiments are terminated. The number of experiments stopping in the unstable stage decreases by 18.27\%. Notably, many experiments in this stage are terminated due to operational issues, such as misconfigured parameters, randomization failures, or deployment bugs, and this stage accounts for a substantial share of all experiments. Although the percentage reduction may appear modest, it is particularly meaningful in practice. In contrast, the numbers of experiments stopping in the overlapping and representative stages increase by 50.80\% and 30.72\%, respectively.
Taken together, these results indicate that adopting our staged framework materially shifts stopping behavior at the platform level toward more representative and reliable experimental outcomes.

\section{Practical Guidelines}
\label{sec:practicalGuide}

In this section, we provide insights on how to apply our framework in real-world settings, including selecting covariate variables to satisfy the prerequisites and outlining the detailed procedure for implementing the approach in practical experiments.

\subsection{Covariate Selection}
\label{sec:covariateSelection}

One challenge is selecting an efficient set of covariates to satisfy both Assumption~\ref{asp:ignorability} and Assumption~\ref{asp:exchangeability}. In online experiments, Assumption~\ref{asp:ignorability} is typically satisfied and can be easily verified through the randomization checks, such as Sample Ratio Mismatch (SRM) test or AA test. Therefore, in the following discussion, we focus solely on the variable selection in accordance with Assumption~\ref{asp:exchangeability}.


To some extent, Assumption~\ref{asp:exchangeability} is quite similar to the unconfoundedness assumption commonly used in observational studies, with the key distinction being that the treatment status is replaced by the indicator of participation in the experiment. In the context of unconfoundedness, we aim to control the effect of treatment assignment on the specific realization of $Y$. However, participation itself does not intrinsically affect the value of the primary outcome $Y$. 
Covariates can influence the estimation of treatment effects when participants and non-participants differ in their distributions and when the treatment exhibits heterogeneous effects across groups.

\begin{figure}[h]
\centering
\includegraphics[width=1\textwidth]{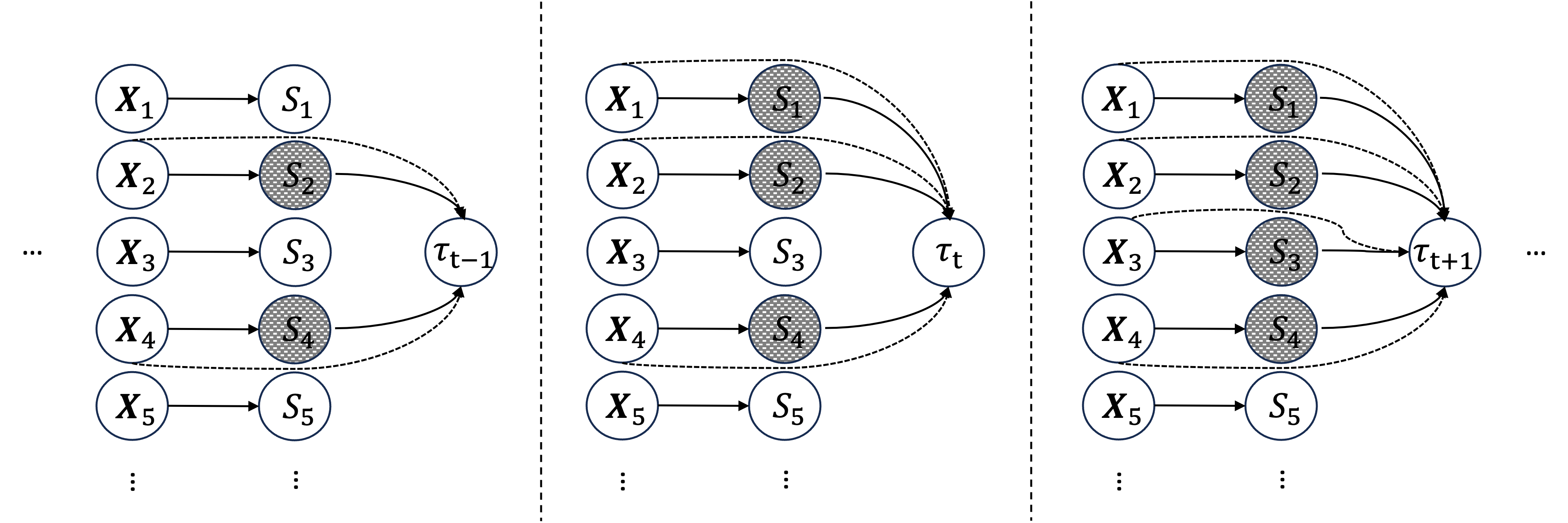}
\caption{Illustration of the relationship between SATE $\tau$, participation indicator $S$, and covariate variables $\bm{X}$ at different time periods.} 
{\footnotesize \textit{Note}: 
Each graph separated by dash line illustrate the condition at a specific time period. Nodes highlighted in bold gray represent their presence during that time period. Solid arrows represent causal paths, indicating the sample population on which the SATE is defined at each time point. Dashed arrows imply indirect paths influencing the SATE, capturing differences in treatment effects across heterogeneous groups.}
\label{fig:causal_diagram}
\end{figure}

Figure~\ref{fig:causal_diagram} further illustrates the relationship between the sample average treatment effect $\tau$, participation $\bm{S}$, and covariate variables $\bm{X}$ over time. We can observe that the sample population is influenced by covariate variables $\bm{X}$, which control the ``weight" of the treatment effect under specific covariate conditions. Simultaneously, the discrepancy between the heterogeneous treatment effect and the average treatment effect is also regulated by covariate variables. Together, these factors lead to fluctuations in the sample average treatment effect over time.


Based on the above analysis, covariates that influence both the heterogeneous treatment effect and the participation indicator $S$ should be considered to select. Furthermore, since the estimation bias - quantified by $\rho$ in Proposition \ref{prop:UpperBound} - relies on the upper bound of the sum of absolute values of the difference-in-means estimator across subgroups (a constant times the absolute value of the difference-in-means estimator), a quantitative approach is to select the (univariate) covariate $\bm{X}$ that maximizes the ratio of $|\hat{\tau}_{HTE}(t, \bm{x})|$ to $|\widehat{\tau}_{DIM}(t)|$. The higher the ratio, the greater the heterogeneity and the potential magnitude of bias. This guideline aligns with the intuition that covariates associated with both treatment effect heterogeneity and the outcome of interest are critical and should be closely monitored.

\subsection{Practical Procedure}
\label{sec:procedure}

Overall, considering the generalizability of the experiment, our method provides additional information for experimenters with diverse objectives, enabling them to make optimal decisions at any stage of the experiment. 
From the experimenters' perspective, we summarize the procedure of our method in Algorithm~\ref{alg:procedure} as follows.

\newcommand{\algorithmicbreak}{\textbf{break}}
\begin{algorithm}[h]
\caption{Procedure for Enhancing External Validity in Online Experiments}\label{alg:method}
\label{alg:procedure}
\begin{algorithmic}[1]
\Require Thresholds $\eta_o$ and $\eta_r$ reflecting the desired trade-off between generalizability and experiment duration
\Require An online experiment with ongoing sampling, where subject arrival times are uncertain over an extended time horizon
\While{the experiment is running at time $t$}
    \Statex \hspace{\algorithmicindent} \textit{Stage detection}
    \State Fit a survival model using each subject's arrival status and observed duration up to time $t$
    \State Compute the heuristic function $\hat{\pi}^{\inf}(t)$ from the fitted survival model
    \Statex
    \Statex \hspace{\algorithmicindent} \textit{Stage classification and estimation}
    \If{$\hat{\pi}^{\inf}(t) \leq \eta_o$} \Comment{Unstable stage}
        \State Reliable inference is not yet available; continue the experiment
    \ElsIf{$\eta_o < \hat{\pi}^{\inf}(t) \leq \eta_r$} \Comment{Overlapping stage}
        \If{the experimenter decides to terminate}
            \State Estimate treatment effect using the IPW estimator $\widehat{\tau}_{IPW}(t)$
            \State \algorithmicbreak
        \EndIf
    \Else \Comment{Representative stage: $\hat{\pi}^{\inf}(t) > \eta_r$}
        \If{the experimenter decides to terminate}
            \State Estimate treatment effect using the DIM estimator $\widehat{\tau}_{DIM}(t)$
            \State \algorithmicbreak
        \EndIf
    \EndIf
\EndWhile
\end{algorithmic}
\end{algorithm}

\newpage

\section{Conclusion}
\label{sec:conclusion}

This paper proposes a phased estimation framework that improves the external validity of online experiments conducted under ongoing sampling. By segmenting the sampling process into three stages---unstable, overlapping, and representative---using a survival-analysis-based heuristic, our framework provides stage-specific estimators for the population average treatment effect, thereby enhancing the transparency of sample representativeness as well as the reliability of experimental results and decision-making.

Our empirical evaluation provides strong evidence of the framework's practical value at scale. In a real-world A/B test on WeChat, the proposed method recovers the true treatment effect as early as the overlapping stage---two days before the unadjusted estimator converges. Applied to 600 platform experiments, the framework increases the true positive rate by 28--37\% while simultaneously reducing the false positive rate by 17--29\%. Furthermore, following the platform-wide deployment of the framework within WeChat's experimentation system, we observe a meaningful shift in experimenters' behavior: average experiment duration increases by 20.7\%, and a significantly higher proportion of experiments conclude in the overlapping and representative stages rather than the unstable stage. These findings suggest that providing real-time diagnostics of sample representativeness, together with improved causal estimation, leads to more reliable experimentation practices.

From a managerial perspective, our framework offers a transparent and operational tool for monitoring the generalizability of experimental results. Rather than imposing a single optimal stopping rule, it provides flexible guidance that accommodates different stakeholder priorities. The framework also serves as a governance mechanism by identifying unstable stages in which reliable inference is not yet available, helping platforms and decision-makers avoid the premature termination of experiments and the costly consequences of biased estimates and suboptimal product deployment decisions.


Several limitations of our work point to promising directions for future research. First, our framework relies on the conditional exchangeability assumption (Assumption 2), which requires that participation timing is independent of potential outcomes given observed covariates. When unobserved factors jointly influence both participation and treatment response, residual bias may persist. Developing formal sensitivity analyses to assess the robustness of the framework under violations of this assumption is an important next step \citep{andrews2017weighting, nguyen2017sensitivity}. Second, the survival models employed here---Kaplan--Meier and Cox proportional hazards---may be restrictive for the high-dimensional, sparse covariate spaces common on digital platforms. More flexible modeling approaches, including deep learning methods for survival analysis \citep{lee2018deephit, lee2019dynamic, ranganath2016deep}, warrant investigation. Third, while we provide guidance for selecting the stage-determining thresholds $\eta_o$ and $\eta_r$, more systematic, data-driven procedures---potentially informed by historical experimentation records---would further reduce the reliance on experimenter judgment. Finally, extending the framework to settings with network interference, where the stable unit treatment value assumption may not hold, represents a natural and important generalization for platforms with strong social interactions among users.

\bibliographystyle{informs2014} 
\bibliography{bibliography} 



\clearpage

\begin{APPENDICES}
\renewcommand\thefigure{\thesection\arabic{figure}}    
\renewcommand\thetable{\thesection\arabic{table}}    
\setcounter{figure}{0}    
\setcounter{table}{0}

\section{Additional Illustration of Proposition~\ref{prop:UpperBound}}

\subsection{Missing Proofs of Proposition~\ref{prop:UpperBound}}
\label{sec:appendix:proof1}
\proof{Proof of Proposition~\ref{prop:UpperBound}.}
From the definition of $\widehat{\tau}_{DIM}(t)$ and $\widehat{\tau}_{IPW}(t)$, we can rewrite them into following terms.
\begin{align*}
\widehat{\tau}_{DIM}(t) &= \frac{2}{\sum_{i = 1}^N \bone\{S_{it} = 1\}} \left(\sum_{i = 1}^N \bone\{S_{it}=1,W_{i}=1\}Y_i(1) - \sum_{i = 1}^N \bone\{S_{it}=1,W_{i}=0\}Y_i(0)\right) \\
&= \sum_{\bm{x}\in\bbX}\frac{\sum_{i=1}^N\bone\{S_{it}=1,\bm{X}_i = \bm{x}\}}{\sum_{i=1}^N\bone\{S_{it}=1\}}\cdot\frac{2}{\sum_{i=1}^N\bone\{S_{it}=1,\bm{X}_i = \bm{x}\}}\cdot \\
&\quad \left(\sum_{i = 1}^N \bone\{S_{it}=1,\bm{X}_i = \bm{x},W_{i}=1\}Y_i(1)
- \sum_{i = 1}^N \bone\{S_{it}=1,\bm{X}_i =\bm{x},W_{i}=0\}Y_i(0)\right) \\
&= \sum_{\bm{x}\in\bbX}2\cdot\frac{\sum_{i=1}^N\bone\{S_{it}=1,\bm{X}_i = \bm{x}\}}{\sum_{i=1}^N\bone\{S_{it}=1\}}\cdot\hat{\tau}_{HTE}(t,\bm{x}).
\end{align*}


\begin{align*}
\widehat{\tau}_{IPW}(t) &= \frac{2}{N} \left(\sum_{i = 1}^N \frac{\bone\{S_{it}=1,W_{i}=1\}}{\hat{\pi}(t\vert\bm{X}_i)}Y_i(1) - \sum_{i = 1}^N \frac{\bone\{S_{it}=1,W_{i}=1\}}{\hat{\pi}(t\vert\bm{X}_i)}Y_i(0)\right) \\
&= \sum_{\bm{x}\in\bbX}\frac{1}{\hat{\pi}(t\vert\bm{X}_i=\bm{x})}\cdot\frac{\sum_{i=1}^N\bone\{S_{it}=1,\bm{X}_i = \bm{x}\}}{N}\cdot\frac{2}{\sum_{i=1}^N\bone\{S_{it}=1,\bm{X}_i = \bm{x}\}}\cdot \\
&\quad \left(\sum_{i = 1}^N \bone\{S_{it}=1,\bm{X}_i = \bm{x},W_{i}=1\}Y_i(1)
- \sum_{i = 1}^N \bone\{S_{it}=1,\bm{X}_i =\bm{x},W_{i}=0\}Y_i(0)\right) \\
&= \sum_{\bm{x}\in\bbX}\frac{2\cdot\sum_{i=1}^N\bone\{S_{it}=1\}}{N\cdot\hat{\pi}(t\vert\bm{X}_i=\bm{x})}\cdot\frac{\sum_{i=1}^N\bone\{S_{it}=1,\bm{X}_i = \bm{x}\}}{\sum_{i=1}^N\bone\{S_{it}=1\}}\cdot\hat{\tau}_{HTE}(t,\bm{x}).
\end{align*}

Note that $\sum_{i = 1}^N\bone\{S_{it} = 1,\bm{X}_i = \bm{x},W_{i}=0\} = \sum_{i = 1}^N\bone\{S_{it} = 1,\bm{X}_i = \bm{x},W_{i}=1\} = \frac{1}{2}\cdot\sum_{i = 1}^N\bone\{S_{it} = 1,\bm{X}_i = \bm{x}\}$ and $\sum_{i = 1}^N\bone\{S_{it} = 1,W_{i}=0\} = \sum_{i = 1}^N\bone\{S_{it} = 1,W_{i}=1\} = \frac{1}{2}\cdot\sum_{i = 1}^N\bone\{S_{it} = 1\}$ holds by default as we assume that the intervention is completely randomly assigned across heterogenous groups defined by $\bm{X}$.

Therefore the estimated absolute difference can be written as
\begin{align*}
|\widehat{\tau}_{DIM}(t) - \widehat{\tau}_{IPW}(t)| 
&= 2\cdot\left|\sum_{\bm{x}\in\bbX}\left(1 - \frac{\sum_{i=1}^N\bone\{S_{it}=1\}}{N\cdot\hat{\pi}(t\vert\bm{X}_i=\bm{x})}\right) \frac{\sum_{i=1}^N\bone\{S_{it}=1,\bm{X}_i = \bm{x}\}}{\sum_{i=1}^N\bone\{S_{it}=1\}}\cdot\hat{\tau}_{HTE}(t,\bm{x})\right|\\
&\leq 2\cdot\left(\frac{1}{\hat{\pi}_{inf}(t)}-1\right) \sum_{\bm{x}\in\bbX}\frac{\sum_{i=1}^N\bone\{S_{it}=1,\bm{X}_i = \bm{x}\}}{\sum_{i=1}^N\bone\{S_{it}=1\}}\cdot\left|\hat{\tau}_{HTE}(t,\bm{x})\right|.
\end{align*}

Since 
$$\sum_{\bm{x}\in\bbX}\frac{\sum_{i=1}^N\bone\{S_{it_j}=1,\bm{X}_i = \bm{x}\}}{\sum_{i=1}^N\bone\{S_{it_j}=1\}}\cdot\left|\hat{\tau}_{HTE}(t,\bm{x})\right|\leq \frac{C}{2}\cdot |\widehat{\tau}_{DIM}(t)|,$$

We can further scale 
the inequality as follows
\begin{align*}
|\widehat{\tau}_{DIM}(t) - \widehat{\tau}_{IPW}(t)| \leq C\cdot|\widehat{\tau}_{DIM}(t)|\cdot\left(\frac{1}{\hat{\pi}_{inf}(t)}-1\right)<\rho
\end{align*}

By the definition of $T_r$, our goal is to identify a specific time $t$ at which the upper bound of the estimated absolute difference falls below the threshold $\rho$. The heuristic function $\hat{\pi}_{inf}(t)$ thus should fulfill the following condition.
$$\hat{\pi}_{inf}(t)>\frac{1}{1+\frac{\rho}{C\cdot |\widehat{\tau}_{DIM}(t)|}}>\frac{C\cdot |\widehat{\tau}_{DIM}(t)|}{\rho + C\cdot |\widehat{\tau}_{DIM}(t)|}$$

\subsection{The Value of $C$}
\label{sec:appendix:C}

Empirically, the constant $C$ can be learned through the historical records. Figure~\ref{fig:distn_c_over_duration} visualizes the values of the constant $C$ in 268 sampled experiments on Weixin experimentation platform. Our empirical result suggests that the value $C$ is relatively stable across the experiments, regardless the duration. Thus, choosing the empirical values of $C=2$ should be relatively robust. Note that in our sampled experiments, the outcome of is interest is some measurement of activeness and the covariates used is active level of participants, whose distribution is changing over the time and is highly correlated to the outcome of interest. 


\begin{figure}[h]
    \centering
    \includegraphics[width=0.5\textwidth]{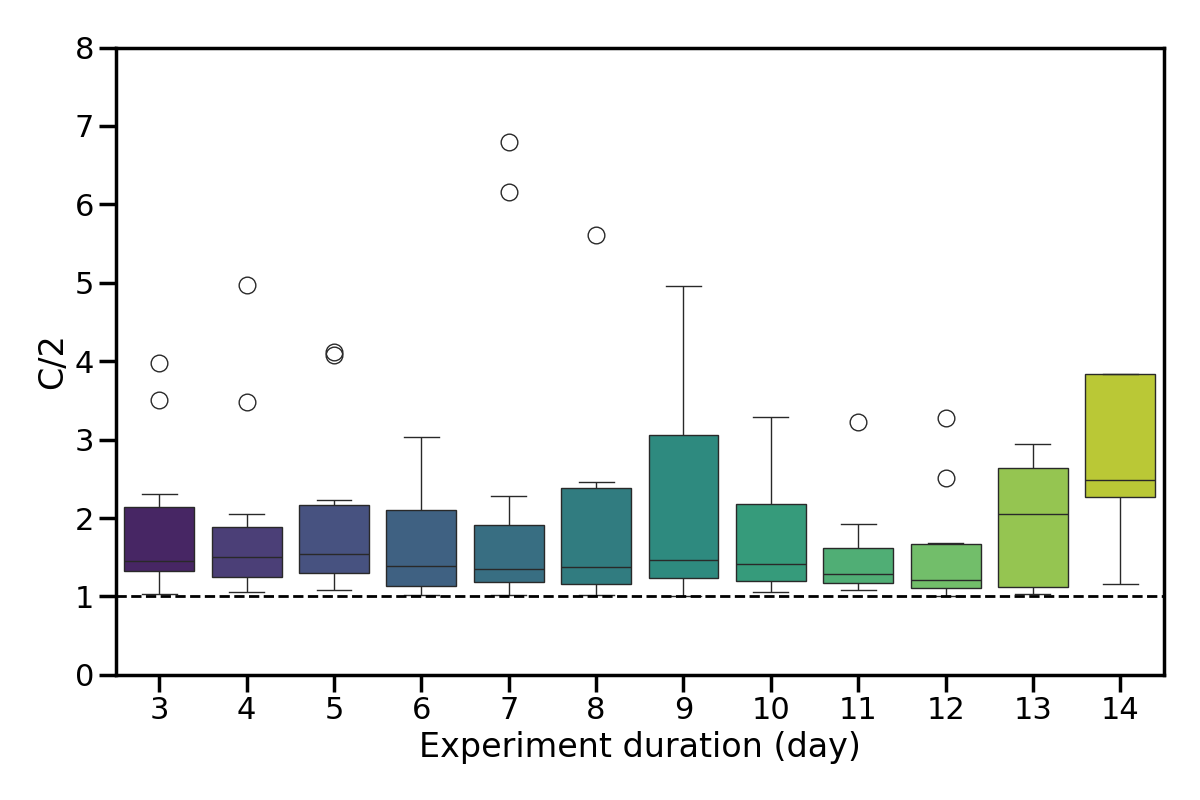} 
    \caption{The boxplot of ratio constant $\frac{C}{2}$ in 268 sampled experiments.}
    \label{fig:distn_c_over_duration}
\end{figure}

\section{A Synthetic Experiment}
\label{sec:simulation}

We conduct a synthetic experiment using accessible individual-level data, as real-world experiments are limited to aggregate-level data exposure due to data privacy concerns.
We simulate an experiment comprising 2000 units for 30-day time period. Each unit is endowed with a covariate variable $X$, which is assigned a random value with equal probability from a sequence of increasing integers started from zero: $\{0,1,2,3\}$. The likelihood of units engaging with the product and being recruited to the experiment at time $t$ is influenced by both the active level dedicated by $X$ and the weekday/weekend effect conditional on $X$. Specifically, let $\vert X\vert$ represent the cardinality of $X$, then $\pi(t\vert X)$, the probability of participation in the experiment at time $t$ given $X$, derives from the following distribution \footnote{Note that here we naturally assume the potential treatment in experiments does not influence the probability of users' exposure to the experiment.}:


\[
\pi(t\vert X)=\left\{
            \begin{array}{ll}
              \mathcal{U}(\frac{x}{|X|}, \frac{x+1}{|X|}),\ \quad\quad if\ t\ is\ on\ weekends\\
              \vspace{0.005em}\\
              \mathcal{U}(\frac{x}{|X|}, \frac{x+1}{|X|})/(x+1),\quad if\ t\ is\ on\ weekdays\\
            \end{array}
          \right.
\]
where $\mathcal{U}(\cdot,\cdot)$ denotes the uniform distribution. 

Before any treatment is initiated, the outcome variable for all units, denoted as $Y$, follows the same normal distribution: $\mathcal{N}(1,0.01)$. We randomly assigned 1000 units to the treatment groups and the other 1000 units to the control groups, and $Y$ for each unit becomes
\[
Y(x)=\left\{
            \begin{array}{ll}
              \mathcal{N}(0.5,0.01) + \mathcal{U}(\frac{x}{|X|}, \frac{x+1}{|X|}),\quad if\ treated\\
              \vspace{0.005em}\\
              \mathcal{N}(1,0.01),\quad\qquad\qquad\qquad if\ controlled\\
            \end{array}
          \right.
\]
We can easily observe that the treatment effect is more pronounced for units for the heterogenous group with a larger $X$, while the ground truth $\tau=0$ indicates no average treatment effect for the population. If we ignore the potential sample bias and directly use the difference-in-mean estimator to estimate the treatment effect, we will almost surely overestimate the true effect if we terminate the experiment in one week, as shown in Figure~\ref{fig:DIM_simul}.

\begin{figure}[ht]
\centering
\includegraphics[width=0.4\textwidth]{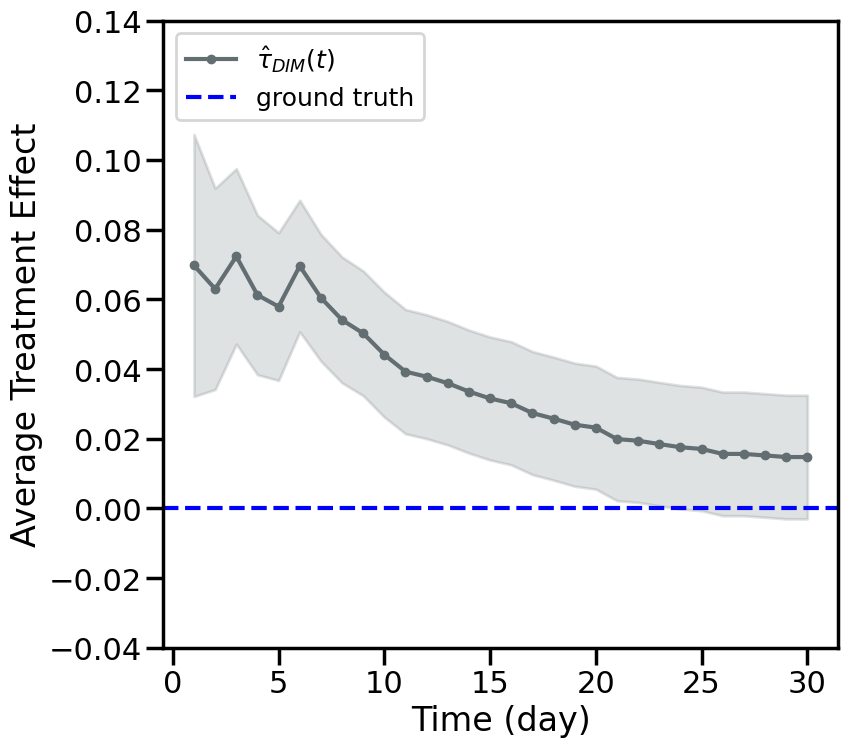}
\caption{Treatment effect estimated by the Difference-in-Means estimator}
\label{fig:DIM_simul}
\end{figure}

We attempt to model the probability of participation of units in different types, i.e. $\pi(t\vert X)$, through the Kaplan-Meier model based on the synthetic data in the pre-treatment period. With the assistance of the `lifeline' toolbox in Python, we can estimate and present $\hat{\pi}(t\vert X)$ in Figure~\ref{fig:Survival_simul}.

\begin{figure}[h]
\centering
\includegraphics[width=0.4\textwidth]{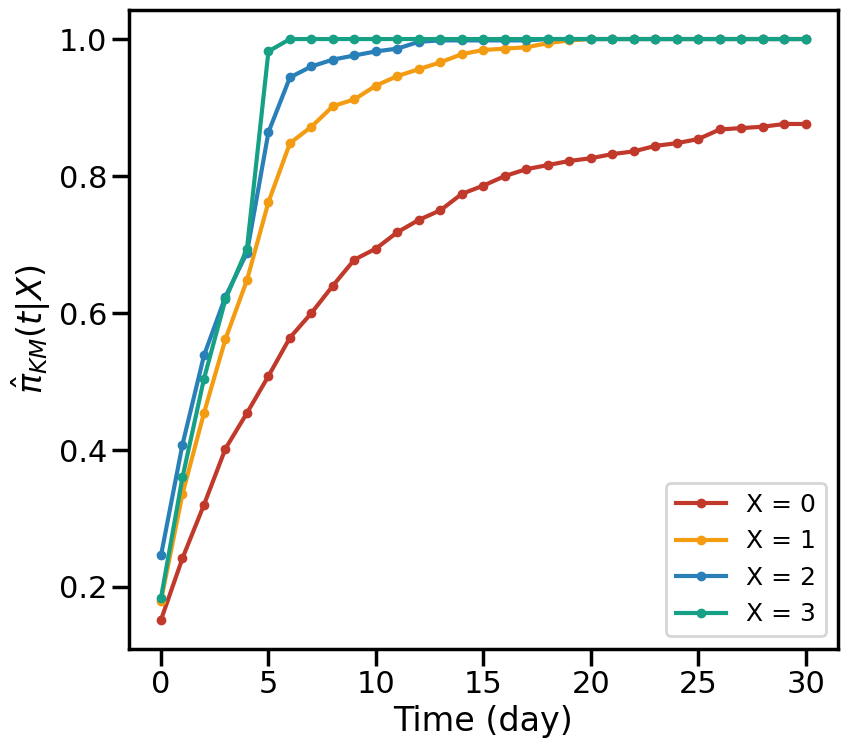}
\caption{The estimation of $\pi(t\vert X)$ generated by the Kaplan-Meier model on the specific value of covariate $X$.}
\label{fig:Survival_simul}
\end{figure}


With the availability of $\hat{\pi}(t|X)$, the computation of the heuristic function $\hat{\pi}^{inf}(t)$ becomes straightforward, enabling the deduction of $T_o$ and $T_r$ with some pre-determined $\eta_o$ and $\eta_r$. 
Figure~\ref{fig:Heuristic_simul} illustrates an example with $\eta_o=0.5$ and $\eta_r=0.85$, showcasing $T_o = 5$ and $T_r = 25$ determined by $\hat{\pi}^{inf}(t)$ generated with the Kaplan-Meier model. 


\begin{figure}[h]
\centering
\includegraphics[width=0.4\textwidth]{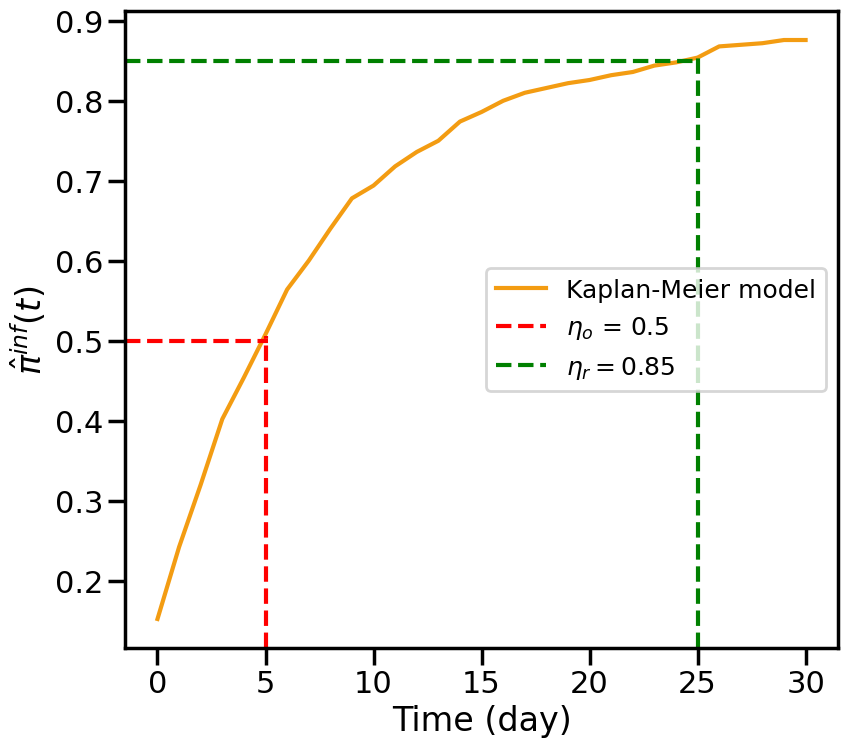}
\caption{The heuristic function $\hat{\pi}^{inf}(t)$ over time. }
{\footnotesize \textit{Note}: 
Yellow solid curve presents the heuristic function generated with the Kaplan-Meier model. Red dash line indicates the point in time when $\hat{\pi}^{inf}(t)$ reaches the threshold $\eta_o=0.5$, which occurs on day 5 for both models. Green dash line indicates the point in time when $\hat{\pi}^{inf}(t)$ reaches the threshold $\eta_r=0.85$, which is $T_r=25$.}
\label{fig:Heuristic_simul}
\end{figure}

Next, we proceed to apply $\hat{\pi}_{KM}(t|X)$ to generate the estimator $\hat{\tau}_{IPW}(t)$, which serves as the debiased estimate. The confidence intervals are derived from the 2.5th and 97.5th percentiles of the 1,000-times bootstrapped estimates. 
Figure~\ref{fig:est_simul} comprehensively presents the performance of the debiased estimator across different stages. Before $T_o$, the estimator exhibits extreme instability, indicating that experiments should not be halted at this stage. For experiments halted after $T_o$, adjusted estimator should be adopted for effective average treatment effect estimation. 

\begin{figure}[h]
\centering
\includegraphics[width=0.4\textwidth]{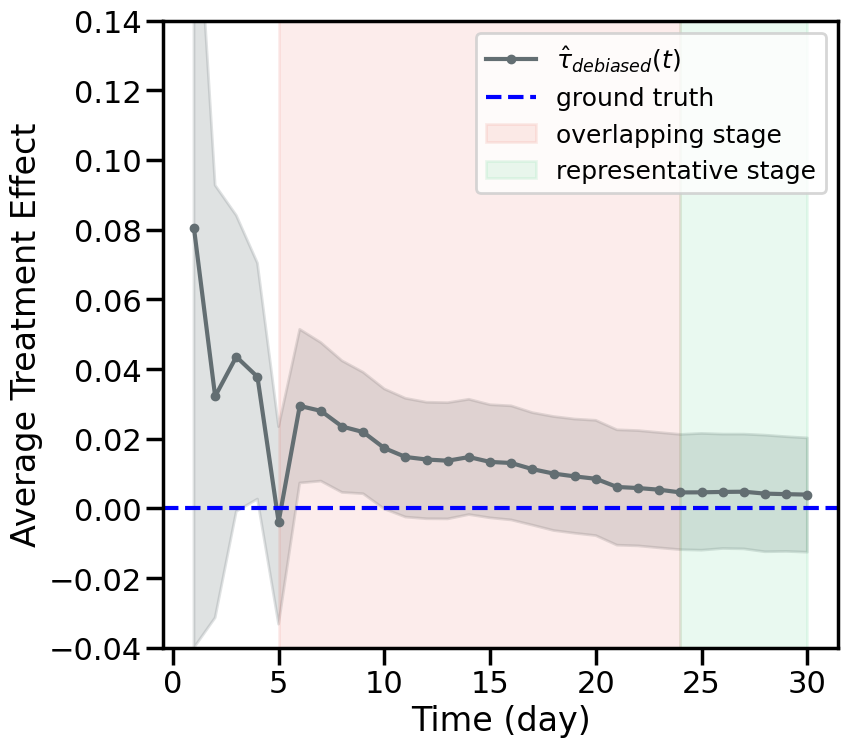}
\caption{Debiased estimation of the average treatment effect at different experiment stages.}
{\footnotesize \textit{Note}: 
Grey solid curves present the debiased estimated effects with shadows indicating 95\% confidence intervals. Blue dash line presents the true average treatment effect. The red area indicates the time interval of the overlapping stage, while the green area indicates the time interval of the representative stage.}
\label{fig:est_simul}
\end{figure}

We further assess the effectiveness of the estimators by analyzing bias and mean squared errors (MSE). For comparison, we include the Jackknife re-sampling estimator proposed by \cite{wang2019heavy}, which, to the best of our knowledge, is the only biased-adjustment estimator that considers external validity in A/B tests. The average bias and MSE over the overlapping stage and the representative stage (time after day $5$) are presented in Table~\ref{tb:MSE_simul}. The results show that the IPW estimator outperforms all methods, including the Difference-in-Means estimator (which serves as the baseline), in terms of MSE. It exhibits slightly weaker performance than the Jackknife re-sampling estimator in terms of bias. This discrepancy is attributed to the relative instability of the Jackknife re-sampling estimator in the early stages, which improves as the duration of the experiment progresses (See Figure~\ref{fig:jackknife_simul} in Appendix).

In conclusion, our stage division criteria are validated, and the effectiveness of the IPW estimator constructed with the survival model is demonstrated through a synthetic experiment.



\begin{table}[h]
\centering
\caption{Comparison result between different estimators in terms of Bias and MSE for the synthetic data}
  \begin{tabular}{c c c}
    \toprule
    {\textbf{Method}} & {\textbf{Bias}} & {\textbf{MSE}} \\
      \midrule
      IPW estimator  & $1.136\times 10^{-2}$ & $2.637\times 10^{-4}$ \\
      Difference-in-Means estimator  & $3.130\times 10^{-2}$ & $1.312\times 10^{-3}$ \\
      Jackknife re-sampling estimator & $7.959\times 10^{-3}$ & $3.363\times 10^{-4}$ \\
    \bottomrule
  \end{tabular}
\label{tb:MSE_simul}
\end{table}

\section{Estimation Methods}

\subsection{IPW Estimator Unbiasedness Proof}
\label{sec:appendix:unbiasProof}

Similar to \cite{stuart2011use}, we provide evidence to show that the expectation of the IPW estimator is equal to the population average treatment effect.

Let $e_{wt}(\bm{X}_i) = Pr\bb{W_i=w\vert \bm{X}_i, S_{it}=1}$ denotes the probability of treatment assignment for participated units, which is similar to the propensity score used in observational study \citep{rosenbaum1983central}. We assume the strong ignorability, $W_i \independent \{Y_i(1), Y_i(0)\}\vert \bm{X}_i$, which is naturally satisfied in randomized controlled experiments. It is obvious that

$$\pi(t\vert\bm{X}_i)\cdot e_{wt}(\bm{X}_i) = Pr\bb{S_{it}=1, W_i=w\vert \bm{X}_i} = \bbE \bc{\bone\{S_{it} = 1,W_{i}=w\}\vert \bm{X}_i}$$


Therefore, we have

\begin{align*}
\bbE\bc{\frac{\bone\{S_{it} = 1,W_{i}=w\}Y_i}{\pi(t\vert\bm{X}_i)e_{wt}(\bm{X}_i)}}  &= \bbE\bc{\frac{\bone\{S_{it} = 1,W_{i}=w\}Y_i(w)}{\pi(t\vert\bm{X}_i)e_{wt}(\bm{X}_i)}} \\ &= \bbE \bb{\bbE \bc{\frac{\bone\{S_{it} = 1,W_{i}=w\}Y_i(w)}{\pi(t\vert\bm{X}_i)e_{wt}(\bm{X}_i)} \Bigg| \bm{X}_i}} \\ 
&= \bbE \bb{ \frac{1}{\pi(t\vert\bm{X}_i)e_{wt}(\bm{X}_i)} \bbE \bc{\bone\{S_{it} = 1,W_{i}=w\} Y_i(w)\vert \bm{X}_i}} \\
&= \bbE \bb{ \frac{1}{\pi(t\vert\bm{X}_i)e_{wt}(\bm{X}_i)} \bbE \bc{\bone\{S_{it} = 1,W_{i}=w\}\vert \bm{X}_i} \bbE \bc{ Y_i(w)\vert \bm{X}_i}}\\ 
&= \bbE \bb{\bbE \bc{Y_i(w)\vert \bm{X}_i}} \\
&= \bbE \bb{Y_i(w)}
\end{align*} 

Due to the randomized controlled setting, $e_{wt}(\bm{X}_i)$ is pre-determined before the experiment starts. A naive estimator to $e_{wt}(\bm{X}_i)$ is $\sum_{i = 1}^N \bone\{S_{it} = 1,W_{i}=w\} /N$. Since the sample mean is an unbiased estimation of expectation under the central limit theorem, the unbiasedness of the IPW estimator is evidently established.

\subsection{Alternative Estimators}
\label{sec:appendix:estimators}
In addition to the IPW estimator discussed in the main context, there are other adjusted estimators commonly employed to correct covariate distribution imbalances. We propose the following estimators as practical alternatives to the IPW estimator.

\subsubsection{Outcome Model-based Estimator}
This approach is suggested under the premise that we can regard the outcome of units that have not participated in the experiment as 'missing data', and fill in with proper values generated by an outcome regression model.
\begin{align*}
\widehat{\tau}(t) = \frac{1}{N}\sum_{i=1}^{N}(\hat{\bbE}[Y_i\vert \bm{X}_i,S_{it}=1,W_i=1] - \hat{\bbE}[Y_i\vert \bm{X}_i,S_{it}=1,W_i=0])
\end{align*}

\subsubsection{Doubly Robust Estimator}
Let $\hat{g}_{wt}(\bm{X}_i) = \hat{\bbE}[Y_i\vert \bm{X}_i,S_{it}=1,W_i=w]$ be the outcome model we used to predict the 'missing' outcomes. By integrating the reweighting approach with the outcome model-based approach, we formulate the following doubly robust estimator.
\begin{multline*}
\widehat{\tau}(t) = \frac{1}{\sum_{i = 1}^N \bone\{S_{it} = 1,W_{i}=1\}}\sum_{i = 1}^N \frac{\bone\{S_{it}=1,W_{i}=1\}}{\hat{\pi}(t\vert\bm{X}_i)}(Y_i-\hat{g}_{1t}) \\
 - \frac{1}{\sum_{i = 1}^N \bone\{S_{it} = 1,W_{i}=0\}}\sum_{i = 1}^N \frac{\bone\{S_{it}=1,W_{i}=0\}}{\hat{\pi}(t\vert\bm{X}_i)}(Y_i-\hat{g}_{0t})
 + \frac{1}{N}\sum_{i=1}^{N}(g_{1t}(\bm{X}_i) - \hat{g}_{0t}(\bm{X}_i)).
\end{multline*}

\subsubsection{Jackknife re-sampling estimator}

\cite{wang2019heavy} introduced a bias-adjusted estimator based on jackknife considering the first-order heavy-user bias in A/B test. We apply this estimator as a comparison to our methods in the main context. The performance of the Jackknife re-sampling estimator in the synthetic experiment is illustrated in Figure~\ref{fig:jackknife_simul}.

\begin{figure}[h]
\centering
\includegraphics[width=0.4\textwidth]{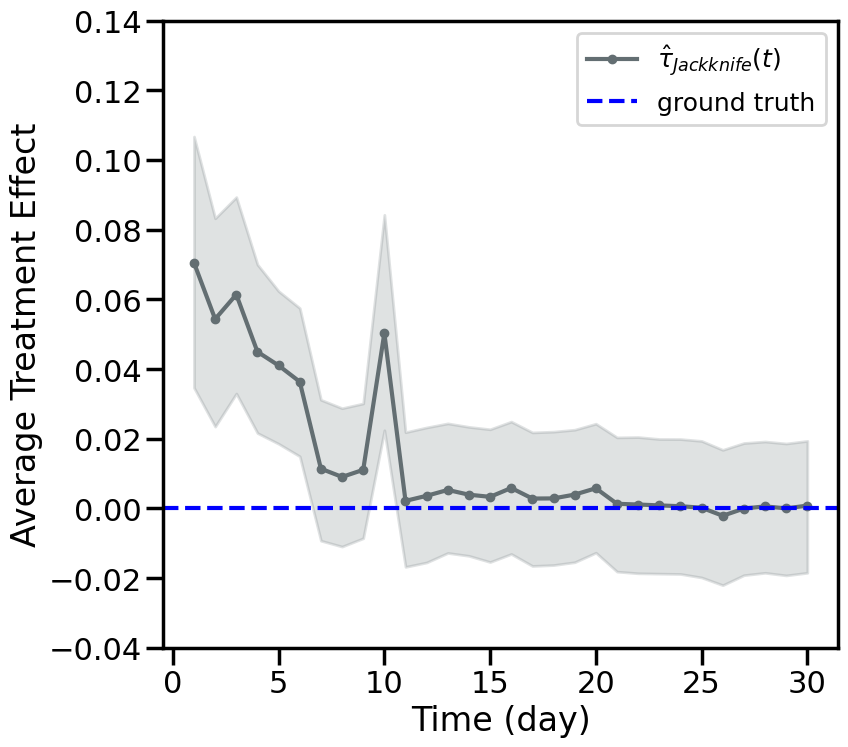}
\caption{Treatment effect estimated by the Jackknife re-sampling estimator in the synthetic experiment.}
\label{fig:jackknife_simul}
\end{figure}




\section{Additional Illustration of the Survival Model}

\subsection{Alternative Survival Model: Cox Proportional Hazards Model}
\label{sec:appendix:cox}

The Cox Proportional Hazards model is a widely used semiparametric survival model, typically employed to estimate the relative hazard—the change in the instantaneous rate of events between groups defined by distinct covariate levels. In our application, we primarily utilize the survival function obtained from a fitted Cox model, from which we derive the estimate of \(\pi(t \mid \bm{X}_i)\):
\[
\hat{\pi}_{CPH}(t \mid \bm{X}_i) = 1 - \hat{S}_0(t)^{\exp(\bm{\hat{\beta}} \cdot \bm{X}_i)},
\]
where \(\hat{S}_0(t)\) denotes the baseline survival function corresponding to the covariate vector \(\bm{X} = \bm{0}\), and \(\bm{\hat{\beta}}\) is the estimated coefficient vector.

The Cox model is one of the most widely used survival models, particularly in clinical research. However, it relies on the critical assumption that the relative hazard remains constant over time across different levels of the covariates \citep{kuitunen2021testing}. When this proportional hazards assumption is violated, alternative strategies can be employed. For example, one may stratify the analysis based on the covariates that do not satisfy the assumption or use an extended Cox model \citep{lin2002modeling}.

\subsection{The Performance of the Survival Model}

We assess the performance of two survival models, the Kaplan-Meier model and the Cox model, using the AUC score. Specifically, we split the entire experiment dataset into 90\% for training and 10\% for testing. Note that at time $t$, we can observe $t$-period experimental data which is right censored, since subjects who have not participated in the experiment are still survived, albeit with an unknown survival duration. We fit the survival model on the training data, assuming that the experiment stops at each time $t$ within 30-day periods, and compute the AUC score on the test data. The results are presented in Figure~\ref{fig:AUC_simul}. The mean AUC score over time for the Kaplan-Meier model is 0.8270 (s.e. 0.009911), while the mean AUC score for the Cox model is 0.8267 (s.e. 0.009880).

\begin{figure}[h]
\centering
\includegraphics[width=0.4\textwidth]{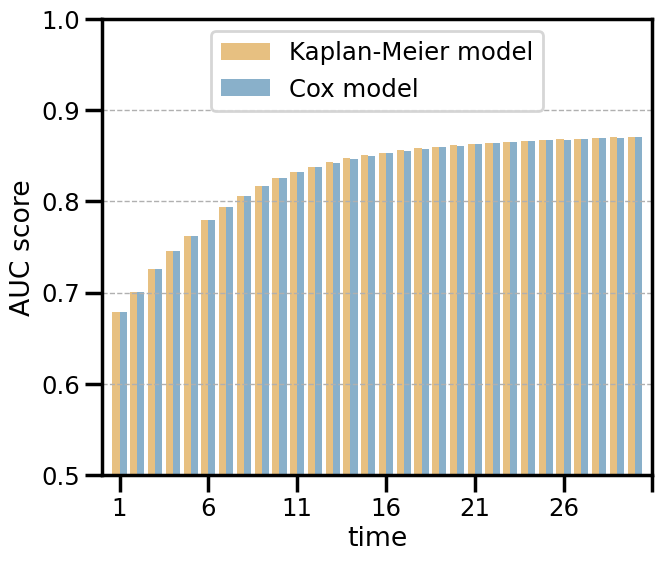}
\caption{The AUC score of survival models at different experiment stopping times.}
\label{fig:AUC_simul}
\end{figure}

\section{Supplement to Experimental Results}

As we showcase the FPR and TPR in Table~\ref{tb:example2} to illustrate the performance of our framework in practice, here we add the original number of experiments for the four metrics (TN, FP, FN, TP) as complementary data.

\begin{table}[h]
\centering
\caption{The number of experiments for each metric among 600 experiments}
\begin{tabular}{c|cccc}
\toprule
& the baseline  & overlapping  & representative & representative \\ 
& approach & ($\eta= 0.5$) & ($\eta' = 0.8$) & ($\eta' = 0.9$)\\
\midrule
True Negative (TN)       & 452               & 462           & 465       & 469         \\
False Positive (FP)     & 58               & 48              & 45          & 41            \\
False Negative (FN)       & 58              & 49              & 46           & 46           \\
True Positive (TP)   & 32                & 41            & 44           & 44     \\ 
\bottomrule
\end{tabular}
\label{tab:example2}
\end{table}

\section{Internal Validity Check}\label{sec:appendix:validity_check}
We assess the internal validity of the running example experiment through two complementary checks: a Sample Ratio Mismatch (SRM) test and an A/A test on pre-treatment covariates.

\noindent\textbf{SRM Test.} Sample Ratio Mismatch (SRM) occurs when the observed number of users in each group does not align with the allocation specified by the experimenters, which is intended to be 1:1. A chi-square test is conducted separately for each of the 9 experiment days to determine whether the traffic allocation meets the expected distribution. After applying the Bonferroni, Benjamini-Hochberg (BH), or Benjamini-Yekutieli (BY) corrections to account for multiple testing\citep{bonferroni1936teoria,hochberg1990more,benjamini1995controlling}, we found no evidence of SRM issues throughout the experiments, either at the significance level of
$\alpha=0.05$ or $\alpha = 0.01$. Figure~\ref{fig:samp_size} confirms that the cumulative sample sizes for the treatment and control groups track each other closely throughout the experiment.

\begin{figure}[h]
\centering
\includegraphics[width=0.4\textwidth]{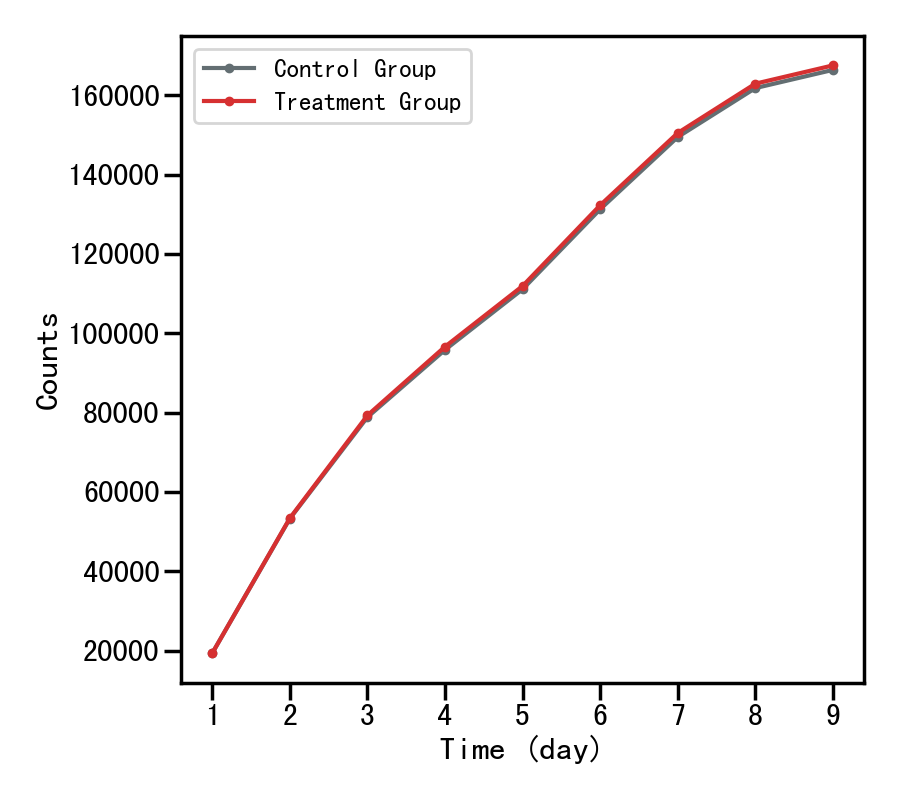}
\caption{Cumulative sample size for treatment and control groups in the experiment}
\label{fig:samp_size}
\end{figure}


\noindent\textbf{A/A Test.} Furthermore, we sought to assess significant differences between the two groups by conducting t-tests on eight key covariates, which were anticipated to show no significant differences. Table~\ref{tab:aa result} presents the results of these tests, comparing the performance of the two groups across the eight key metrics during the seven days leading up to the start of the experimentation. These eight metrics include the primary outcome of interest, one activeness metric, and six consumption metrics, selected by the experimenters based on their domain expertise. As users accumulate on a daily basis, we conducted the same tests for each day. After applying the Bonferroni, Benjamini-Hochberg (BH), or Benjamini-Yekutieli (BY) procedures to account for multiple hypothesis testing, we found that none of the key metrics exhibited statistically significant differences on any day at the $\alpha = 0.05$ or $\alpha = 0.01$ levels. This indicates an internally valid traffic allocation for comparable experimental groups. In total, $8 \times 9 = 72$ hypothesis tests were conducted (8 metrics across 9 days). Even without multiple testing corrections, the raw p-values are overwhelmingly insignificant, and no test reaches significance at $\alpha = 0.05$ after applying any of the three correction procedures.
\begin{table}[h]
\centering
\caption{Relative differences of pre-treatment outcomes for 8 key metrics (p-values in parentheses) }\scriptsize
\begin{tabular}{llllllllll}
\hline
Metric & Day 1                                                        & Day 2                                                        & Day 3                                                        & Day 4                                                        & Day 5                                                        & Day 6                                                        & Day 7                                                        & Day 8                                                        & Day 9                                                        \\ \hline
Primary     & \begin{tabular}[c]{@{}l@{}}0.419\%\\ (0.45769)\end{tabular}  & \begin{tabular}[c]{@{}l@{}}0.663\%\\ (0.04822)\end{tabular}  & \begin{tabular}[c]{@{}l@{}}0.581\%\\ (0.04204)\end{tabular}  & \begin{tabular}[c]{@{}l@{}}0.493\%\\ (0.05641)\end{tabular}  & \begin{tabular}[c]{@{}l@{}}0.363\%\\ (0.13)\end{tabular}     & \begin{tabular}[c]{@{}l@{}}0.344\%\\ (0.11762)\end{tabular}  & \begin{tabular}[c]{@{}l@{}}0.323\%\\ (0.11935)\end{tabular}  & \begin{tabular}[c]{@{}l@{}}0.3\%\\ (0.13247)\end{tabular}    & \begin{tabular}[c]{@{}l@{}}0.292\%\\ (0.13908)\end{tabular} 
 \\
Activeness      & \begin{tabular}[c]{@{}l@{}}1.259\%\\ (0.38484)\end{tabular}  & \begin{tabular}[c]{@{}l@{}}0.563\%\\ (0.52598)\end{tabular}  & \begin{tabular}[c]{@{}l@{}}0.557\%\\ (0.45338)\end{tabular}  & \begin{tabular}[c]{@{}l@{}}0.405\%\\ (0.55428)\end{tabular}  & \begin{tabular}[c]{@{}l@{}}0.312\%\\ (0.62634)\end{tabular}  & \begin{tabular}[c]{@{}l@{}}0.354\%\\ (0.55271)\end{tabular}  & \begin{tabular}[c]{@{}l@{}}0.223\%\\ (0.6921)\end{tabular}   & \begin{tabular}[c]{@{}l@{}}0.134\%\\ (0.80574)\end{tabular}  & \begin{tabular}[c]{@{}l@{}}0.076\%\\ (0.8888)\end{tabular}   \\
Consump. 1      & \begin{tabular}[c]{@{}l@{}}5.033\%\\ (0.2321)\end{tabular}   & \begin{tabular}[c]{@{}l@{}}1.216\%\\ (0.57619)\end{tabular}  & \begin{tabular}[c]{@{}l@{}}2.127\%\\ (0.23415)\end{tabular}  & \begin{tabular}[c]{@{}l@{}}2.744\%\\ (0.12604)\end{tabular}  & \begin{tabular}[c]{@{}l@{}}3.083\%\\ (0.0665)\end{tabular}   & \begin{tabular}[c]{@{}l@{}}3.913\%\\ (0.01172)\end{tabular}  & \begin{tabular}[c]{@{}l@{}}3.474\%\\ (0.01778)\end{tabular}  & \begin{tabular}[c]{@{}l@{}}3.179\%\\ (0.02476)\end{tabular}  & \begin{tabular}[c]{@{}l@{}}2.951\%\\ (0.03436)\end{tabular}  \\
Consump. 2      & \begin{tabular}[c]{@{}l@{}}2.172\%\\ (0.17794)\end{tabular}  & \begin{tabular}[c]{@{}l@{}}0.624\%\\ (0.51075)\end{tabular}  & \begin{tabular}[c]{@{}l@{}}0.648\%\\ (0.42173)\end{tabular}  & \begin{tabular}[c]{@{}l@{}}0.581\%\\ (0.44114)\end{tabular}  & \begin{tabular}[c]{@{}l@{}}0.596\%\\ (0.40142)\end{tabular}  & \begin{tabular}[c]{@{}l@{}}0.882\%\\ (0.18027)\end{tabular}  & \begin{tabular}[c]{@{}l@{}}0.744\%\\ (0.23381)\end{tabular}  & \begin{tabular}[c]{@{}l@{}}0.615\%\\ (0.30905)\end{tabular}  & \begin{tabular}[c]{@{}l@{}}0.514\%\\ (0.38927)\end{tabular}  \\
Consump. 3      & \begin{tabular}[c]{@{}l@{}}-0.676\%\\ (0.68749)\end{tabular} & \begin{tabular}[c]{@{}l@{}}-0.572\%\\ (0.55406)\end{tabular} & \begin{tabular}[c]{@{}l@{}}-0.519\%\\ (0.51384)\end{tabular} & \begin{tabular}[c]{@{}l@{}}-0.547\%\\ (0.45269)\end{tabular} & \begin{tabular}[c]{@{}l@{}}-0.584\%\\ (0.39193)\end{tabular} & \begin{tabular}[c]{@{}l@{}}-0.289\%\\ (0.64583)\end{tabular} & \begin{tabular}[c]{@{}l@{}}-0.251\%\\ (0.67314)\end{tabular} & \begin{tabular}[c]{@{}l@{}}-0.268\%\\ (0.64082)\end{tabular} & \begin{tabular}[c]{@{}l@{}}-0.329\%\\ (0.56286)\end{tabular} \\
Consump. 4      & \begin{tabular}[c]{@{}l@{}}-0.761\%\\ (0.6112)\end{tabular}  & \begin{tabular}[c]{@{}l@{}}-0.286\%\\ (0.7392)\end{tabular}  & \begin{tabular}[c]{@{}l@{}}-0.312\%\\ (0.65693)\end{tabular} & \begin{tabular}[c]{@{}l@{}}-0.34\%\\ (0.59852)\end{tabular}  & \begin{tabular}[c]{@{}l@{}}-0.475\%\\ (0.43397)\end{tabular} & \begin{tabular}[c]{@{}l@{}}-0.34\%\\ (0.54317)\end{tabular}  & \begin{tabular}[c]{@{}l@{}}-0.369\%\\ (0.48524)\end{tabular} & \begin{tabular}[c]{@{}l@{}}-0.418\%\\ (0.41306)\end{tabular} & \begin{tabular}[c]{@{}l@{}}-0.468\%\\ (0.35402)\end{tabular} \\
Consump. 5      & \begin{tabular}[c]{@{}l@{}}-0.761\%\\ (0.6112)\end{tabular}  & \begin{tabular}[c]{@{}l@{}}-0.286\%\\ (0.7392)\end{tabular}  & \begin{tabular}[c]{@{}l@{}}-0.312\%\\ (0.65693)\end{tabular} & \begin{tabular}[c]{@{}l@{}}-0.34\%\\ (0.59852)\end{tabular}  & \begin{tabular}[c]{@{}l@{}}-0.475\%\\ (0.43397)\end{tabular} & \begin{tabular}[c]{@{}l@{}}-0.34\%\\ (0.54317)\end{tabular}  & \begin{tabular}[c]{@{}l@{}}-0.369\%\\ (0.48524)\end{tabular} & \begin{tabular}[c]{@{}l@{}}-0.418\%\\ (0.41306)\end{tabular} & \begin{tabular}[c]{@{}l@{}}-0.468\%\\ (0.35402)\end{tabular} \\
Consump. 6      & \begin{tabular}[c]{@{}l@{}}0.474\%\\ (0.74931)\end{tabular}  & \begin{tabular}[c]{@{}l@{}}0.36\%\\ (0.67702)\end{tabular}   & \begin{tabular}[c]{@{}l@{}}0.226\%\\ (0.75208)\end{tabular}  & \begin{tabular}[c]{@{}l@{}}0.156\%\\ (0.81267)\end{tabular}  & \begin{tabular}[c]{@{}l@{}}0.015\%\\ (0.98103)\end{tabular}  & \begin{tabular}[c]{@{}l@{}}0.217\%\\ (0.70528)\end{tabular}  & \begin{tabular}[c]{@{}l@{}}0.193\%\\ (0.72119)\end{tabular}  & \begin{tabular}[c]{@{}l@{}}0.123\%\\ (0.81438)\end{tabular}  & \begin{tabular}[c]{@{}l@{}}0.048\%\\ (0.92532)\end{tabular}  \\ \hline
\end{tabular}
\label{tab:aa result}
\end{table}

\section{Sensitivity Analysis}\label{sec:appendix:sensitivity_analysis}

We evaluate the performance of our framework-based estimators in the simulation settings described in Section~\ref{sec:simulation} using two metrics: bias and mean squared error (MSE). Table~\ref{tab:sensitivityCheck} reports results for two alternative survival models, the Kaplan-Meier estimator and the Cox proportional hazards model, across different values of \(\eta_o\) and \(\eta_r\). Overall, when \(\eta_o < 0.5\), the estimates remain relatively stable across parameter settings, indicating that our approach is robust to moderate variation in \(\eta_o\).

The results also reveal a clear bias-variance tradeoff governed by the threshold parameters. For the overlapping stage, increasing \(\eta_o\) from 0.35 to 0.50 substantially reduces bias (e.g., from \(4.354 \times 10^{-2}\) to \(3.827 \times 10^{-3}\) for KM) at the cost of a later \(T_o\) and thus a longer experiment duration. However, MSE also decreases, indicating that tighter overlap requirements yield more precise estimates overall. For the representative stage, a similar pattern holds: increasing \(\eta_r\) from 0.75 to 0.90 reduces bias at the expense of longer waiting times (\(T_r\) increases from 14 to 29 days for KM). Notably, the MSE for the DIM estimator at the representative stage is consistently lower than that of the IPW estimator at the overlapping stage, confirming the practical value of waiting for representativeness when feasible. These results provide practitioners with concrete guidance: more conservative (higher) thresholds reduce estimation bias but require longer experiments, and the choice should reflect the experimenter's tolerance for bias relative to the cost of delayed decisions.

\begin{table}[ht]
\centering
\caption{Sensitivity analysis with varying values of $\eta_o$ and $\eta_r$ for debiased estimation across different survival models.}
\renewcommand{\arraystretch}{1.2}
\begin{tabular}{c c cc cc cc}
\hline
 & & \multicolumn{2}{c}{$T_o$} & \multicolumn{2}{c}{bias} & \multicolumn{2}{c}{MSE} \\
\cline{3-4} \cline{5-6} \cline{7-8}
 &  & KM & CPH & KM & CPH & KM & CPH \\
\hline
$\eta_o = 0.35$ & & 3 & 4 & $4.354\times10^{-2}$ & $2.888\times10^{-2}$ & $2.356\times10^{-3}$ & $1.332\times10^{-3}$ \\
$\eta_o = 0.40$ & & 3 & 5 & $4.354\times10^{-2}$ & $1.568\times10^{-4}$ & $2.356\times10^{-3}$ & $1.931\times10^{-4}$ \\
$\eta_o = 0.45$ & & 4 & 5 & $3.778\times10^{-2}$ & $1.568\times10^{-4}$ & $1.725\times10^{-3}$ & $1.931\times10^{-4}$ \\
$\eta_o = 0.50$ & & 5 & 5 & $3.827\times10^{-3}$ & $1.568\times10^{-4}$ & $2.254\times10^{-4}$ & $1.931\times10^{-4}$ \\
\hline
$\eta_r = 0.75$ & & 14 & 11 & $1.476\times10^{-2}$ & $1.727\times10^{-2}$ & $2.862\times10^{-4}$ & $3.620\times10^{-4}$ \\
$\eta_r = 0.80$ & & 17 & 14 & $1.129\times10^{-2}$ & $1.742\times10^{-2}$ & $1.957\times10^{-4}$ & $3.658\times10^{-4}$ \\
$\eta_r = 0.85$ & & 25 & 18 & $4.635\times10^{-3}$ & $1.272\times10^{-2}$ & $8.309\times10^{-5}$ & $2.263\times10^{-4}$ \\
$\eta_r = 0.90$ & & 29 & 27 & $4.112\times10^{-3}$ & $6.542\times10^{-3}$ & $8.713\times10^{-5}$ & $1.122\times10^{-4}$ \\
\hline
\end{tabular}
\label{tab:sensitivityCheck}
\end{table}

\end{APPENDICES}




\end{document}